\documentclass{aa}  

\usepackage{graphicx}
\usepackage{txfonts}
\usepackage{amsmath}	
\usepackage{amssymb}	
\usepackage{xcolor}
\usepackage[]{hyperref}
\usepackage[]{natbib}

\newcommand{\civ}{\ion{C}{IV}}	
\newcommand{\mgii}{\ion{Mg}{II}}
\newcommand{\siv}{\ion{Si}{IV}}	
\newcommand{\lya}{Ly$\alpha$}	
\newcommand{\hi}{\ion{H}{I}}
\newcommand{\maggiv}{MAGG IV}
\newcommand{\maggv}{MAGG V}

\begin{document}

   \title{MUSE Analysis of Gas around Galaxies (MAGG) - VI.  The cool and enriched gas environment of $z\gtrsim 3$ Ly$\alpha$ emitters.}

    \titlerunning{The cool and enriched gas environment of $z\gtrsim 3$ Ly$\alpha$ emitters.}
    \authorrunning{M. Galbiati et al.} 

   \author{
          Marta Galbiati\inst{1}\fnmsep\thanks{m.galbiati@campus.unimib.it},
          Rajeshwari Dutta\inst{2},
          Michele Fumagalli\inst{1,3}\fnmsep\thanks{michele.fumagalli@unimib.it},
          Matteo Fossati\inst{1,4},
          \and
          Sebastiano Cantalupo\inst{1}
          }

   \institute{Dipartimento di Fisica G. Occhialini, Universit\`a degli Studi di Milano Bicocca, Piazza della Scienza 3, 20126 Milano, Italy
        \and
            IUCAA, Postbag 4, Ganeshkind, Pune 411007, India
        \and
             INAF – Osservatorio Astronomico di Trieste, via G. B. Tiepolo 11, I-34143 Trieste, Italy
        \and 
            INAF - Osservatorio Astronomico di Brera, via Bianchi 46, 23087 Merate (LC), Italy\\     
             }

   \date{\today}

 

\abstract{We present a novel dataset that extends our view of the cosmic gas around $z\approx 3-4$ Ly$\alpha$ emitting galaxies (LAEs) in the Muse Analysis of Gas around Galaxies (MAGG) survey by tracing a cool and enriched gas phase through 47 \mgii\ absorbers identified in newly-obtained VLT/XSHOOTER near-infrared quasar spectra. Jointly with the more ionized gas traced by \civ\ systems and the neutral \hi\ from previous work, we find that LAEs are distributed inside cosmic structures that contain multiphase gas in composition and temperature. 
All gas phases are a strong function of the large-scale galaxy environment: the \mgii\ and the \civ\ strength and kinematics positively correlate with the number of associated galaxies, and it is $\approx3-4$ times more likely to detect metal absorbers around group than isolated LAEs. Exploring the redshift evolution, the covering factor of \mgii\ around group and isolated LAEs remains approximately constant from $z\approx3-4$ to $z\lesssim2$, but the one of \civ\ around group galaxies drops by $z\lesssim2$. Adding the cool enriched gas traced by the \mgii\ absorbers to the results we obtained for the \hi\ and \civ\ gas, we put forward a picture in which LAEs lie along gas filaments that contain high column-density \hi\ systems and are enriched by strong \civ\ and \mgii\ absorbers. While the \mgii\ gas appears to be more centrally concentrated near LAEs, weaker \civ\ systems trace instead a more diffuse gas phase extended up to larger distances around the galaxies.}


\maketitle
%

\section{Introduction}

The gas reservoir retained within the intergalactic medium (IGM) and circumgalactic medium (CGM) stands as a critical regulator in the formation, growth, and evolution of galaxies. While metal-poor gas flows through filaments connecting the galaxies and provides fuel for sustaining their star formation \citep{Keres2005, Dekel2009}, the balance between accretion and feedback processes enriches and shapes the properties of the CGM gas, rendering it multiphase in temperature, density, kinematics, and composition \citep{Oppenheimer2008,Lilly2013,Tumlinson2017}.
A complete view of the connection and the co-evolution of galaxies with the surrounding cosmic gas is thus crucial to paint a comprehensive picture of how galaxies form and evolve through time.
In addition to the galaxies' internal processes, the large-scale galaxy environment is expected to affect the activity of the galaxies themselves and the properties of their surrounding gas via hydrodynamic and gravitational interactions \citep[see e.g.][]{Gunn1972,Merritt1983,Boselli2006}. 

Targeting the diffuse cosmic gas in absorption in the spectra of background bright sources, such as quasars, has been widely adopted as a powerful technique to connect the multiphase CGM and (often massive star-forming) galaxies across cosmic time (see \citealp{Chen2017} for a review at $z<0.5$ and \citealp{Steidel2010,Rudie2012,Turner2014,Rudie2019} at $z\approx2-3$). Doublet transitions of metals, such as \mgii\ $\lambda\lambda 2796,2803$ and \civ\ $\lambda\lambda 1548,1550$, are ideal probes of the different gas phases ranging from the cool ($T\approx10^4\rm\,K$) and enriched gas to a more ionized and, sometimes, warmer phase ($T\approx10^5\rm\,K$). At the same time, large spectroscopic surveys, especially with modern integral field spectrographs (IFS) that allow the assembly of dense and complete samples of galaxies, have been instrumental in investigating the gas-galaxy connection down to the lower mass galaxy population. 

\citet{Bergeron1988} and \citet{Bergeron1991} pioneered the observations of \mgii\ absorption systems in quasar sightlines and reported the detection of a large number of galaxies at the redshift of the absorbers. Since then, the \mgii\ has been widely used as an effective tracer of cool gas around galaxies across cosmic time (see \citealp{Barnes2014} for a review), even up to the reionization epoch with recent James Webb Space Telescope observations \citep{Bordoloi2023}. Indeed, \mgii-bearing gas is found to be an effective tracer of inflows and outflows \citep{Tremonti2007, Weiner2009, Martin2009, Rubin2010, Nestor2011, Ribaudo2011, Thom2011, Rubin2012, Rubin2014, xu2023}, and to exhibit tight correlations with the galaxies' stellar mass and ongoing star formation activity \citep{Gauthier2009, Chen2010, Bordoloi2011, Menard2011, Churchill2013a, Nielsen2013b_general, Bordoloi2014, Liang2014, Lan2014, Lan2020, Dutta2021}.

A more thorough perspective emerges when different gas phases are observed simultaneously. In the COS-Halos survey \citet{Werk2013,Werk2014} reported that low and medium ionized metals tracing $T\approx10^4-10^5\rm\,K$ gas are common features in the CGM of $L\sim L^\star$ galaxies at $z\approx0.2$. They also found a difference in the radial distribution of different gas phases around galaxies: the gas's ionization state is higher at larger distances from the host, possibly due to the gas being more diffuse and less shielded. Indeed, several results from both simulations \citep{Ford2013,Ho2020,Ho2021,Weng2024} and observations \citep{Schroetter2021,Dutta2020,Qu2023} show that the presence of low-ionization ions (e.g. \mgii, \ion{C}{ii} and \ion{O}{ii}) is confined at closer distances from the galaxies, while higher ions (e.g., \civ, \siv, \ion{O}{iv} and \ion{Ne}{viii}) extend up to several times the galaxies' virial radius. Furthermore, a tomographic analysis of absorption lines performed by \citet{Dutta2024} revealed that the \civ\ gas show higher degree of coherence compared to the \mgii. \citet{Schroetter2021} investigated the connection between \civ\ and \mgii\ absorbers and a population of [\ion{O}{ii}] emitters in the MusE GAs FLow and Wind (MEGAFLOW) survey \citep{Schroetter2016}. They found that if, on the one hand, the size of the \mgii\ halo decreases with time and becomes smaller compared to the dark-matter halo, on the other hand, the \civ\ gas progressively increases in size, possibly indicating a co-evolution with the dark-matter halo. At $z<2$, \citet{Dutta2021} showed that different gas phases do not only vary in their distribution, but also in how they respond to the properties of the galaxies. Indeed, they found that the correlation between the absorption strength and the galaxies' stellar mass and star formation rate is stronger for the \mgii\ than for the \civ. 

Besides individual gas-galaxy connections, an additional dimension must be taken into account to fully trace the co-evolution of gas and galaxies as a function of cosmic time, especially in the context of hierarchical structure formation and interactions in dense galaxy environments. At $z\lesssim1$ several detections of strong \hi\ absorbers connected to multiple galaxies have been reported as tracers of an extended intragroup medium \citep[see e.g.][]{Peroux2017,  Fumagalli2017, Chen2019, Hamanowicz2020}. The cool and enriched phase traced by ions like \mgii\ has also been detected around group galaxies. The presence of multiple galaxies, at $z\lesssim2$, raises the detection rate of the \mgii\ absorption-line systems, compared to what is observed around isolated galaxies \citep{Nielsen2018_grp, Dutta2020, Dutta2021, Qu2023}. These systems are also stronger and show more complex kinematics \citep{Chen2010, Nielsen2018_grp, Fossati2019, Nelson2021}. Both the higher incidence rate of the absorbers and the effects on the gas properties are not fully explained in superposition models, requiring a contribution from environmental interactions that boost the gas cross-section instead \citep[see e.g.][]{Nielsen2018_grp, Dutta2020}. 
At $z\gtrsim3$, when massive structures are not yet collapsed, and environmental interactions are not expected to affect the properties of the surrounding gas as in the local Universe, the picture is less clear. Given the accessibility of \hi\ and \civ\ transitions to optical ground-based spectrographs, the cosmic gas traced by these two ions has been studied as a function of the large scale-scale galaxy environment around massive Lyman Break Galaxies (LBGs; see, e.g., \citealp{Adelberger2005}) and, recently, Ly$\alpha$ Emitters (LAEs). 
By using the Multi Unit Spectroscopic Explorer (MUSE; \citealp{Bacon2010}) at VLT, \citet{Muzahid2021} targeted a population of LAEs at $z\gtrsim3$ and reported that both the \hi\ and \civ\ absorption are stronger in the stacked spectra of groups, relative to isolated LAEs. Similarly, \citet{Lofthouse2023} (\maggiv\ hereafter) and \citet{MAGGV} reported elevated covering fractions of
\hi\ and \civ\ around group galaxies. 
To better understand the origin of the excess absorption in overdensities at $z\approx 3-4$ and the evolution with cosmic time
%
we expand the core dataset of the MUSE Analysis of Galaxy (MAGG) survey \citealp{Lofthouse2019, Dutta2020, Fossati2021, Lofthouse2023, MAGGV} with a new  VLT/XSHOOTER program (PID 0109.A$-$0559; PI M. Galbiati). The MAGG survey is based on a MUSE Large Programme (ID 197.A-0384; PI M. Fumagalli) targeting 28 fields centered on $z\approx3.2-4.5$ quasars with magnitudes $m_{\rm r}\leq 19$ mag and at least one $N_{\rm HI}\ge 10^{17}~\rm cm^{-2}$ hydrogen absorption line system at redshift $z\geq3.05$. By complementing the MUSE observations with archival high-resolution ($ R\gtrsim30,000$) and moderate or high signal-to-noise ($S/N\gtrsim10$ per pixel) spectroscopy of the central quasars, this survey is specifically designed for studying the connection between galaxies and gas across the cosmic time blindly. Leveraging new data, this study focuses on the cool gas traced by 47 \mgii\ absorbers in the near infrared (NIR), complementing the view of more ionized \civ-bearing gas from \citet{MAGGV} (\maggv\ hereafter).


We describe the dataset and the new NIR observations in Section \ref{sec:observations}. In Section \ref{sec:catalogs} we report the details of the samples of \mgii\ absorbers, strong \hi\ Lyman Limit Systems from \cite{Lofthouse2019}, \civ\ systems and LAEs from \maggv\. The correlation between LAEs and the \mgii-bearing gas is explored in Section \ref{sec:MgII-LAE}, and in Section \ref{sec:effect_of_groups} we investigate the effect of the large-scale galaxy environment on the multiphase gas. Finally, in Section \ref{sec:discussion}, we discuss how different gas phases evolve with time in connection to galaxies and the large-scale environment. 
Throughout, unless otherwise noted, we quote magnitudes in the AB system, distances in physical units, and adopt the Planck 2015 cosmology ($\Omega_{\mathrm{m}} = 0.307$, $H_0 = 67.7$ km s$^{-1}$ Mpc$^{-1}$; \citealp{Planck2016}). 


\section{Observations and data reduction}
\label{sec:observations}

\subsection{MUSE observations}

The 28 quasar fields included in the MAGG survey have been observed during ESO periods 97-103 with an integration time of $\approx4$ hours per field using MUSE in the wide field mode with extended wavelength coverage ($465-930$~nm, $R\approx 2000-3500$). All the observations have been carried out on clear nights at airmass $\rm\leq1.6$, resulting in better image quality than $\rm FWHM=0.8\arcsec$. Small dithers and relative rotations of 90 degrees for each exposure have been included to mitigate the effects of sensitivity variations across the $\rm\approx1\times1\,\rm arcmin^{2}$ field of view (FoV). Longer exposures, up to $\approx10$ hours, characterize fields with available archival exposures from GTO observations. All the details, including the exposure times and the image quality for each field, are listed in table 1 of \citet{Lofthouse2019}.

We summarize the MUSE data reduction next, and additional details can be found in \citet{Lofthouse2019}. We first applied bias subtraction, sky flats correction, and flux and wavelength calibrations to the raw data using the ESO MUSE pipeline \citep[][version 2 or greater]{Weilbacher2014}. The datacubes were then post-processed with the {\sc CubExtractor} package ({\sc CubEx}, \citealp{Cantalupo2019}), one of the tools currently available to improve the imperfections arising from the sky subtraction and different response of MUSE spectrographs \citep[see details in][]{Lofthouse2019}. The single reduced exposures are combined into the final products using mean and median statistics. Additional datacubes, containing only half of the exposures each, are produced to facilitate the detection of sources and efficiently identify possible contaminants. We described our method to propagate the uncertainties and estimate the standard deviation in \citet{Lofthouse2019} and \citet{Fossati2019}.

\subsection{Archival optical quasar spectroscopy}

The MAGG survey includes high-resolution ($ R\gtrsim30,000$) and moderate or high signal-to-noise ($S/N\gtrsim10$ per pixel) archival spectroscopy at optical wavelengths for the central quasar of each field. The available observations are obtained with the Ultraviolet and Visual Echelle Spectrograph (UVES, \citealp{Dekker2000}) at the VLT, the High Resolution Echelle Spectrometer (HIRES, \citealp{Vogt1994}) at Keck, the Magellan Inamori Kyocera Echelle (MIKE; \citealp{Bernstein2003}) at the Magellan telescopes. Complementary medium-resolution ($R\gtrsim5,000$) spectra are also available from the Echellette Spectrograph and Imager (ESI, \citealp{Sheinis2002}) at Keck and XSHOOTER \citealp{Vernet2011} at the VLT. A detailed description of the data and the reduction steps can be found in section 3.1 in \citet{Lofthouse2019} with the wavelength coverage, spectral resolution, and signal-to-noise ratio at the selected wavelengths listed in table 2. 

\subsection{XSHOOTER near-infrared spectroscopy}
\label{sec:data_XS}

\begin{figure} 
\centering
\includegraphics[width=\columnwidth]{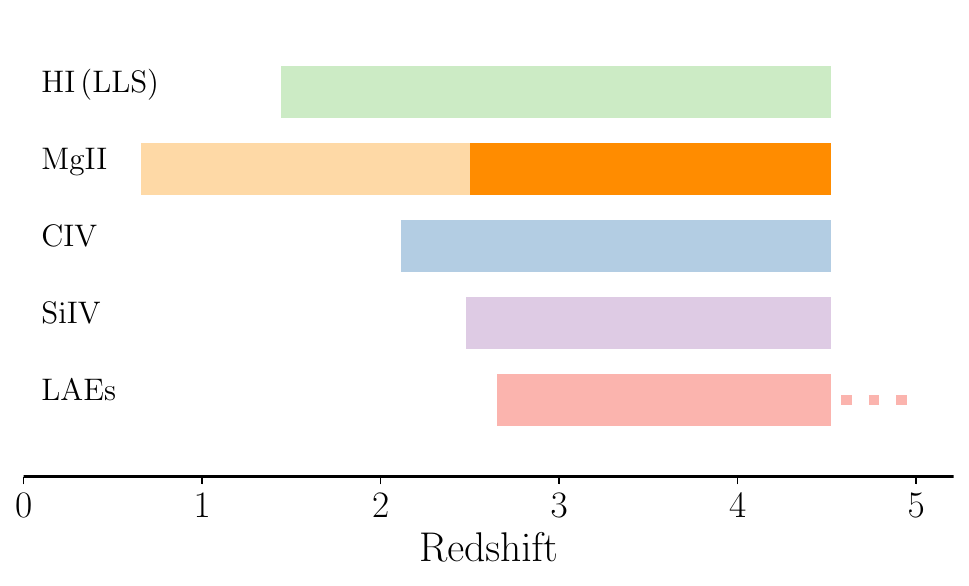}
\caption{Redshift path of the \hi\ (green), \mgii\ (orange), \civ\ (blue) and \siv\ (purple) absorption line systems which trace the multiphase gas around LAEs (red) covered by optical-to-NIR spectroscopy in MAGG and shown up to the redshift of the most distant quasar. The coverage of the \mgii\ at $z\gtrsim3$ (dark orange) is enabled by the new XSHOOTER observations (Section \ref{sec:data_XS}). The redshift range of the absorbers extends from redward the Ly$\alpha$ emission of the lowest redshift quasar up to $3000\rm\,km\,s^{-1}$ from the redshift of the highest quasar of the sample. Telluric regions are not masked in this figure.}
\label{fig:metals_in_MAGG}
\end{figure}

Extending the wavelength coverage up to the near-infrared is essential to tracing the cool gas via \mgii\ absorption at $z>3$. As shown in Figure \ref{fig:metals_in_MAGG}, the updated redshift coverage of the MAGG survey for \hi\ (green), \mgii\ (orange for optical spectroscopy, dark-orange for the extended range probed by NIR spectroscopy), \civ\ (blue) and \siv\ (purple) overlap with that of the low-mass galaxy population traced by LAEs (red), making it feasible to explore different gas phases around the LAEs simultaneously. 

Archival XSHOOTER spectroscopy is available for a subset of MAGG quasars (13 of the 28 quasars) from the data release of the XQ-100 survey \citep{Lopez2016} and, in case of the sightlines J$015741-010629$ and J$020944+051713$, from the ESO archive. We complete NIR observations of the MAGG survey by observing the remaining 15 quasars with a 15.9h XSHOOTER program (PID 0109.A$-$0559; PI M. Galbiati). The observations were executed during ESO periods 109-110 with clear sky, $\leq80\%$ of lunar illumination, and at airmass $\leq1.5$, requiring the image quality to be better than 1.1 arcsec. The nominal resolving power is $\approx5600$ for the 0.9 arcsec slit.

The archival XQ-100 spectra were reduced as described by \citet{Lopez2016} by using a custom IDL pipeline to perform accurate sky subtraction and optimal extraction of the science spectra on unrectified 2D arrays of slit position and wavelength. Each frame is dark subtracted and flat fielded. A b-spline fit \citep{Kelson2003} is used to model the sky emission that is then subtracted. The counts in each 2D frame is flux calibrated using observations of spectro-photometric standard stars taken, whenever possible, close in time to the science observations. As a last step, a final 1D spectrum is extracted for each exposure, rebinned into a common grid, and finally co-added. 

We used the open-source python package {\sc Pypeit} \citep{Prochaska2019, Prochaska2020a} to reduce the new XSHOOTER observations included in MAGG. The key steps of the process are summarized as follows: 
\begin{enumerate}
    \item each raw frame is overscan and bias subtracted; 
    \item the master arc frame is used for wavelength calibration; 
    \item a 2D model of the master flat frame is applied to the science frames and used to trace the echelle orders and correct for the detector illumination; 
    \item cosmic rays are removed using the Laplacian Cosmic Ray Identification algorithm, {\sc l.a. cosmic} \citep{Dokkum2001}; 
    \item the sky emission is subtracted from the 2D frames and modeled by both the AB-pair nodding patterns and iterative b-spline fit to remove residuals; 
    \item the quasar trace is detected automatically and optimally extracted \citep{Horne1986} to generate the 1D and 2D science spectra; 
    \item the flux is calibrated using XSHOOTER standard stars. 
\end{enumerate}   

If multiple exposures for each object are available, these are resampled into a common wavelength grid and then co-added. We used {\sc pypeit} to correct for telluric absorption. As a final step, we modeled the quasar continuum using {\sc ContFit}\footnote{\url{https://github.com/trystynb/ContFit}} which provides a python-based GUI to manually select points in the regions of the quasar spectra that are free of absorption, to be used as knots for a cubic spline.

We report in Table \ref{tab:NIR_spec} the detailed list of the 15 MAGG quasar targeted by the new VLT/XSHOOTER spectra. The Table includes the redshift of each quasar as well as the median signal-to-noise ratios per pixel at the two rest-frame wavelengths $\lambda_{\rm RF}=3000\,\text{\AA}$ and $\lambda_{\rm RF}=3600\,\text{\AA}$ (computed within the wavelength ranges $\Delta\lambda=2990-3010\,\text{\AA}$ and $\Delta\lambda=3590-3610\,\text{\AA}$, respectively), to easily compare the new observations with the archived data of the remaining 13 quasars from \citet{Lopez2016}. 

\begin{table*}
\centering
\begin{tabular}{ccccc}
\hline
Sightline & Redshift & $S/N_{\rm 3000\text{\AA}}^{(a)}$ & $S/N_{\rm 3600\text{\AA}}^{(b)}$ & Wavelength (\AA)$^{(c)}$ \\
\hline
J010619+004823 &  4.4402 &  14  & 12  & 10195-24768 \\
J033413-161205 &  4.3800 &  23  & 17  & 9817-24767  \\
J033900-013318 &  3.2040 &  21  & 18  & 10194-24766 \\
J094932+033531 &  4.1072 &  26  & 28  & 10195-24767 \\ 
J095852+120245 &  3.2746 &  16  & 21  & 10195-24768 \\
J102009+104002 &  3.1528 &  23  & 20  & 10195-24768 \\
J120917+113830 &  3.0836 &  15  & 13  & 10195-24769 \\
J123055-113909 &  3.5570 &  15  & 19  & 10195-24771 \\
J142438+225600 &  3.6340 &  128 & 115 & 10194-24765 \\
J193957-100241 &  3.7870 &  31  & 52  & 9816-24767  \\
J200324-325144 &  3.7850 &  16  & 24  & 10195-24768 \\
J205344-354652 &  3.4900 &  12  & 18  & 9817-24766  \\
J230301-093930 &  3.4774 &  22  & 25  & 10195-24768 \\
J231543+145606 &  3.3971 &  18  & 15  & 10194-24767 \\
J233446-090812 &  3.3261 &  19  & 23  & 10195-24768 \\
\hline
\end{tabular}
\caption{List of the MAGG sightlines targeted by the new VLT/XSHOOTER program PID 0109.A$-$0559. $^{(a)}$ Median signal-to-noise ratio per pixel at rest-frame $\lambda_{\rm RF}=3000\,\text{\AA}$ computed within the range $\Delta\lambda_{\rm RF}=2990-3010\,\text{\AA}$; $^{(b)}$ Median signal-to-noise at $\lambda_{\rm RF}=3600\,\text{\AA}$ within the range $\Delta\lambda_{\rm RF}=3590-3610\,\text{\AA}$; $^{(c)}$ NIR wavelength coverage.}
\label{tab:NIR_spec}
\end{table*}


\section{Data analysis}
\label{sec:catalogs}


\subsection{Catalogue of \mgii\ absorbers}

To trace a cool and enriched gas phase surrounding LAEs, we assemble a sample of $z\gtrsim3$ \mgii\ $\lambda\lambda$ 2796, 2803  absorption line systems identified by visually inspecting the near-infrared quasar spectra available in MAGG. We restrict the search of \mgii\ doublets between the wavelengths redward the quasars' Ly$\alpha$ emission line, being the metals the minority of the transitions in the Ly$\alpha$ forest. We also mask the region at $\Delta v<3000\rm\,km\,s^{-1}$ from the redshift of the quasars (corresponding to observed wavelength range $13153-13285\,$\AA\ at the mean redshift of the quasars included in MAGG, $z_{\rm QSO}^{\rm mean}=3.75$), to avoid proximity effects \citep[see, e.g.,][]{Hennawi2007}. In addition, we mask the wavelength region $13500-14200\,$\AA\ to preserve the sample from any possible contamination due to strong telluric features. Candidates with a strong-to-weak \mgii\ line optical depth ratio inconsistent with the expected value $\approx2:1$ in the unsaturated regime are identified as interlopers and excluded from the following analysis. The final sample includes 47 \mgii\ absorbers at $2.9\lesssim z\lesssim3.8$ (median $z\approx3.31$) and rest-frame equivalent width $W_{\rm r}\gtrsim0.07\,$\AA. 

Following the procedure described in \maggv\, we model the absorption lines identified in the continuum-normalized quasar spectra with Voigt profiles using the Bayesian code MC-ALF \citep{Longobardi2023}. Each system is decomposed into the minimum number of individual Voigt components required to reproduce the observed flux and then convolved with a Gaussian kernel to match the line spread function of the observed data. We assign the following priors to the free parameters of each Voigt component: $1\leq b/(\mathrm{km\,s^{-1}})\leq30$ for the Doppler parameter and $12\leq\log(N/\mathrm{cm^{-2}})\leq16$ for the column density. To minimize the effect of imperfections in the continuum normalization, we include a multiplicative constant ranging between 0.98 and 1.02 as a free parameter. In addition, we have extensively tested (see \maggiv\ and \maggv) that allowing the spectral resolution to vary between $\pm5\rm\,km\,s^{-1}$ around the nominal value significantly improves the fit quality. In the case of line blending, we use additional {\it filler} Voigt components to model the absorption features arising from different ions at different redshifts that are uncorrelated with the targeted absorption-line system. For each \mgii\ system we provide a measure of the rest-frame equivalent width (median $W_{\rm r}\approx 0.58\,$\AA), column density (median $\log(N/{\rm cm^{-2}})\approx 14.3$) and width of the line in velocity space, defined as the velocity interval enclosing the $90\%$ of the absorption line optical depth (median $\Delta v_{90}\approx 110\rm\,km\,s^{-1}$).

\begin{figure} 
\centering
\includegraphics[width=\columnwidth]{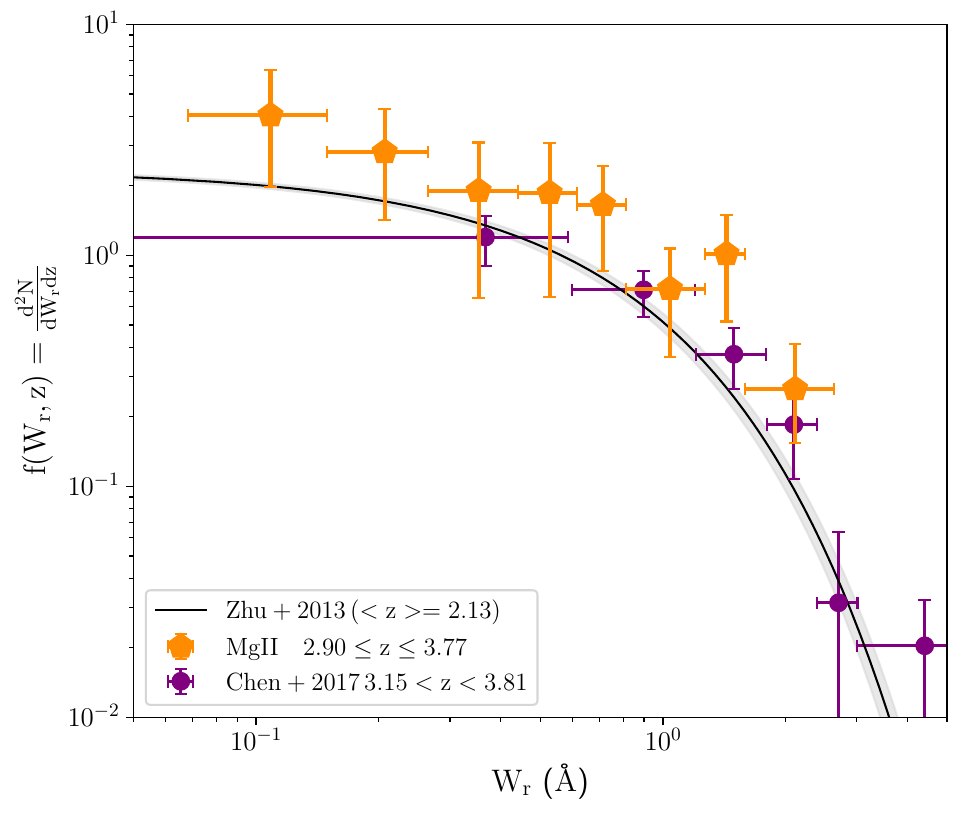}
\caption{Equivalent width distribution function per unit redshift of the MAGG \mgii\ absorption line systems (orange pentagons). The 10th and 90th percentiles from the bootstrap re-sampling and the width of each bin of equivalent width are shown as vertical and horizontal error bars, respectively. The results for the MAGG sample are consistent with the completeness-corrected functions from \citet{Zhu2013} (black solid line) and \citet{Chen2017} (purple dots) down to $W_{\rm r}\approx0.1\,$\AA.}
\label{fig:fwz}
\end{figure}

To assess the completeness of the sample, as done in \maggv, we compute the equivalent width distribution function of the \mgii\ per unit redshift, $f(W_{\rm r},z)=d^2N/dW_{\rm r}/dz$. To quantify the uncertainties, we bootstrap over the sightlines $10^3$ times and derive the 10th and 90th percentiles from the resulting distribution. We compare the results derived for the MAGG \mgii\ sample to the completeness corrected functions estimated for statistical surveys over a similar redshift range in Figure \ref{fig:fwz}. In particular, we refer to the sample of \mgii\ systems found in $\approx10^5$ quasar spectra from SDSS at $z\approx2.05-2.30$ from \citet{Zhu2013} and 75 \mgii\ absorbers identified at $z\approx 3.15-3.81$ in the Magellan/FIRE spectra of 100 quasars from \citet{Chen2017}. The distribution function of the MAGG sample lacks any completeness correction. However, since it does not deviate significantly compared to the expectation from the literature, we conclude that the MAGG \mgii\ sample is reasonably complete down to equivalent width $W_{\rm r}\approx 0.10\,$\AA. The properties of the entire sample of \mgii\ absorbers identified in MAGG are listed in Table \ref{tab:line_list}.

\begin{table*}
\centering
\begin{tabular}{ccccc}
\hline
Sightline & $\mathrm{z^{(a)}}$ & $\mathrm{W_{2796}\,(\AA)}$ & $\mathrm{\log(N_{2796}/cm^{-2})}$ & $\mathrm{\Delta v_{90}\,(km\,s^{-1})}$ \\
\hline
J010619 & 3.7290 & $0.771 \pm 0.031$ & $14.0 \pm 0.2$ & $78.6 \pm 0.1$ \\
J012403 & 2.9865 & $0.906 \pm 0.014$ & $14.8 \pm 0.2$ & $209.1 \pm 9.5$ \\
J012403 & 3.0774 & $0.414 \pm 0.007$ & $13.7 \pm 0.1$ & $76.0 \pm 0.1$ \\
J012403 & 3.6745 & $0.316 \pm 0.011$ & $13.3 \pm 0.3$ & $95.1 \pm 0.1$ \\
J013340 & 3.1394 & $0.152 \pm 0.014$ & $12.6 \pm 0.0$ & $151.7 \pm 9.5$ \\
J013340 & 3.6206 & $0.444 \pm 0.037$ & $15.7 \pm 1.4$ & $113.9 \pm 9.6$ \\
J013340 & 3.6916 & $0.509 \pm 0.005$ & $14.4 \pm 0.3$ & $95.1 \pm 0.1$ \\
J013340 & 3.7732 & $2.635 \pm 0.098$ & $15.5 \pm 0.2$ & $209.0 \pm 19.1$ \\
J013724 & 3.0740 & $0.269 \pm 0.024$ & $13.9 \pm 0.8$ & $75.7 \pm 0.1$ \\
J013724 & 3.1015 & $0.991 \pm 0.089$ & $15.0 \pm 0.6$ & $152.0 \pm 9.5$ \\
J015741 & 3.3855 & $1.500 \pm 0.166$ & $14.8 \pm 0.7$ & $234.6 \pm 22.0$ \\
J020944 & 3.3092 & $0.094 \pm 0.013$ & $12.4 \pm 0.1$ & $44.8 \pm 7.5$ \\
J020944 & 3.6660 & $0.569 \pm 0.086$ & $15.1 \pm 0.4$ & $55.1 \pm 0.0$ \\
J033413 & 2.9267 & $1.549 \pm 0.079$ & $15.3 \pm 0.6$ & $281.0 \pm 9.9$ \\
J033413 & 3.5567 & $0.720 \pm 0.040$ & $14.9 \pm 0.6$ & $82.0 \pm 9.3$ \\
J033900 & 3.0622 & $1.376 \pm 0.060$ & $15.7 \pm 0.2$ & $111.4 \pm 17.1$ \\
J094932 & 3.3100 & $2.112 \pm 0.236$ & $15.5 \pm 0.4$ & $157.2 \pm 17.4$ \\
J111008 & 3.4413 & $0.072 \pm 0.012$ & $12.4 \pm 0.1$ & $76.0 \pm 19.1$ \\
J111113 & 3.4817 & $0.581 \pm 0.006$ & $14.0 \pm 0.1$ & $75.7 \pm 0.2$ \\
J111113 & 3.6075 & $0.575 \pm 0.007$ & $14.8 \pm 0.2$ & $94.8 \pm 10.0$ \\
J111113 & 3.7603 & $0.434 \pm 0.013$ & $13.2 \pm 0.0$ & $209.0 \pm 8.1$ \\
J120917 & 3.0229 & $1.534 \pm 0.042$ & $15.1 \pm 0.6$ & $145.5 \pm 9.0$ \\
J124957 & 3.1023 & $1.239 \pm 0.086$ & $13.9 \pm 0.9$ & $151.8 \pm 25.0$ \\
J133254 & 3.4211 & $0.237 \pm 0.004$ & $13.0 \pm 0.0$ & $75.7 \pm 5.0$ \\
J142438 & 3.5399 & $0.149 \pm 0.005$ & $12.7 \pm 0.0$ & $64.0 \pm 9.0$ \\
J162116 & 3.1057 & $1.587 \pm 0.013$ & $15.4 \pm 0.2$ & $209.0 \pm 9.5$ \\
J193957 & 3.0719 & $0.068 \pm 0.014$ & $12.2 \pm 0.1$ & $69.3 \pm 15.8$ \\
J193957 & 3.2561 & $0.171 \pm 0.008$ & $13.4 \pm 0.2$ & $68.7 \pm 2.0$ \\
J193957 & 3.5724 & $0.149 \pm 0.007$ & $12.9 \pm 0.3$ & $46.7 \pm 8.1$ \\
J200324 & 2.8976 & $0.242 \pm 0.044$ & $12.9 \pm 1.3$ & $78.3 \pm 9.2$ \\
J200324 & 2.9782 & $0.256 \pm 0.022$ & $13.0 \pm 0.4$ & $134.1 \pm 8.3$ \\
J200324 & 3.1724 & $0.145 \pm 0.015$ & $12.8 \pm 0.4$ & $70.0 \pm 9.6$ \\
J200324 & 3.1880 & $2.033 \pm 0.187$ & $15.3 \pm 0.7$ & $354.7 \pm 8.9$ \\
J200324 & 3.3331 & $0.485 \pm 0.033$ & $13.2 \pm 0.3$ & $172.1 \pm 9.8$ \\
J200324 & 3.5500 & $0.648 \pm 0.028$ & $14.4 \pm 0.8$ & $214.4 \pm 9.0$ \\
J205344 & 2.9901 & $1.489 \pm 0.069$ & $15.2 \pm 0.8$ & $350.1 \pm 9.7$ \\
J205344 & 3.0939 & $0.722 \pm 0.013$ & $15.1 \pm 0.7$ & $75.1 \pm 8.3$ \\
J205344 & 3.1720 & $0.969 \pm 0.045$ & $14.1 \pm 0.6$ & $125.8 \pm 9.7$ \\
J221527 & 3.6616 & $0.666 \pm 0.008$ & $14.8 \pm 0.2$ & $76.0 \pm 5.4$ \\
J221527 & 3.7015 & $0.878 \pm 0.023$ & $13.8 \pm 0.0$ & $113.9 \pm 12.8$ \\
J221527 & 3.7091 & $0.143 \pm 0.009$ & $12.8 \pm 0.1$ & $75.7 \pm 9.4$ \\
J231543 & 2.9439 & $1.661 \pm 0.051$ & $15.3 \pm 0.5$ & $208.3 \pm 17.8$ \\
J231543 & 3.2740 & $1.007 \pm 0.035$ & $14.8 \pm 0.9$ & $123.3 \pm 8.3$ \\
J233446 & 3.0569 & $1.674 \pm 0.104$ & $15.6 \pm 0.4$ & $146.8 \pm 10.2$ \\
J234913 & 2.9226 & $0.758 \pm 0.097$ & $14.3 \pm 0.8$ & $209.0 \pm 28.5$ \\
J234913 & 3.6185 & $0.388 \pm 0.023$ & $13.2 \pm 0.0$ & $95.1 \pm 7.6$ \\
J234913 & 3.6922 & $0.404 \pm 0.021$ & $15.6 \pm 0.5$ & $76.0 \pm 9.5$ \\
\hline
\end{tabular}
\caption{List and properties of the \mgii\ absorption-line systems identified in MAGG. {\it (a)} the redshift is based on the observed wavelength of the strongest Voigt component of each line according to the best-fit model.}
\label{tab:line_list}
\end{table*}

\subsection{Catalogues of LAEs, groups and \civ\ absorbers}
\label{sec:catalogs_MAGG}

In \maggv\, we focused on studying how an intermediate ionized gas phase, traced by \civ\ absorption line systems, is linked to a population of low-mass galaxies identified as Ly$\alpha$ emitters in the MUSE cubes. We refer to \maggv\ (section 3 and tables 1-5) for the detailed description of the identification and classification processes of the two samples, providing only the essential information here. Both catalogs are available as supplementary online material to \maggv.

The galaxy sample (``MAGG LAEs'' hereafter) consists of 921 high-confidence Ly$\alpha$ emitters observed at $2.81\lesssim z\lesssim6.60$ (median $z\approx3.91$), with a median Ly$\alpha$ luminosity $\log[L_{\rm Ly\alpha}/({\rm erg\,s^{-1}})]\approx42.10$. This sample is designed to extend the study of the gas-galaxy connection down to typically low stellar and halo masses, of the order of $M_\star\approx10^9-10^{10}\rm\,M_\odot$ and $M_{\rm h}\approx10^{10}-10^{11}\rm M_\odot$, respectively \citep[see e.g.][]{Ouchi2020}. The strategy we used to build this sample follows a two-step approach. For continuum-detected course, we identify rest-frame UV-bright sources by running {\sc sextractor} \citep{Bertin1996} on the white-light images, and we used the output segmentation map to extract the spectra and derive a reliable measure of the redshift using the M. Fossati's fork \footnote{\url{https://matteofox.github.io/Marz}} of {\sc marz} \citep{Hinton2016} for 1200 sources. For line emitters, we first subtracted all the continuum sources in the MUSE cubes and searched for line emission running {\sc CubEx} to identify groups of $\geq27$ connected voxels with signal-to-noise ratio $S/N\geq3$. We then visually inspected all the candidates to avoid contamination from cosmic rays and other emission lines at lower redshift, assigning high confidence to the sources detected at integrated signal-to-noise ratio $ISN\geq7$ and lacking the presence of double peaks with separation consistent with [OII] emitters. The Ly$\alpha$ flux is measured using the curve of growth analysis in pseudo-narrow band images and then corrected for the Milky Way extinction using the map from \citet{Schlafly2011} and the extinction law from \citet{Fitzpatrick1999}. 

To characterize the galaxy environment, we identified 152 groups, broadly defined as $\geq2$ LAEs that are not isolated within a LOS separation of $|\Delta v|<500\rm\,km\,s^{-1}$, without any specific constraint on the halo mass. This evidence suggests that only a fraction of the galaxies in our sample (corresponding to $\approx44$ percent of the LAEs found in MAGG) appear to evolve in isolation. In contrast, for a large number of LAEs that reside in overdense regions, the gravitational and hydrodynamic interactions that take place in dense galaxy environments, as well as the large-scale structure in which groups reside, are found to affect the properties of different CGM gas phases in a large redshift interval (\citealp[see e.g.][]{Fossati2019, Dutta2020, Dutta2021, Muzahid2021, Banerjee2023} and \maggv).

The sample of \civ\ systems identified in MAGG (``MAGG \civ'' hereafter) includes 220 doublets ($\lambda\lambda1548,\,1550\,$\AA) at $3.0\lesssim z\lesssim 4.3$ (median $z\approx3.31$) identified in the high and medium resolution optical spectra available for each quasar of the survey. This sample is $50\%$ complete at rest-frame equivalent widths $W_{\rm r}^{\rm\civ}\approx0.08\,$\AA\ for the strongest component at $\lambda_{\rm RF}=1548\,$\AA. The absorption strength of the systems identified in MAGG spans a broad range of equivalent width, from the weak systems at $W_{\rm r}\lesssim0.08\,$\AA\ up to strongest ones with $W_{\rm r}\gtrsim0.30\,$\AA\ and a median of $W_{\rm r}\approx0.09\,$\AA.

\subsubsection{Overlap of \mgii\ with the \hi\ and \civ\ samples in MAGG}
\label{sec:overlap_MAGG}

\begin{figure} 
\centering
\includegraphics[width=\columnwidth]{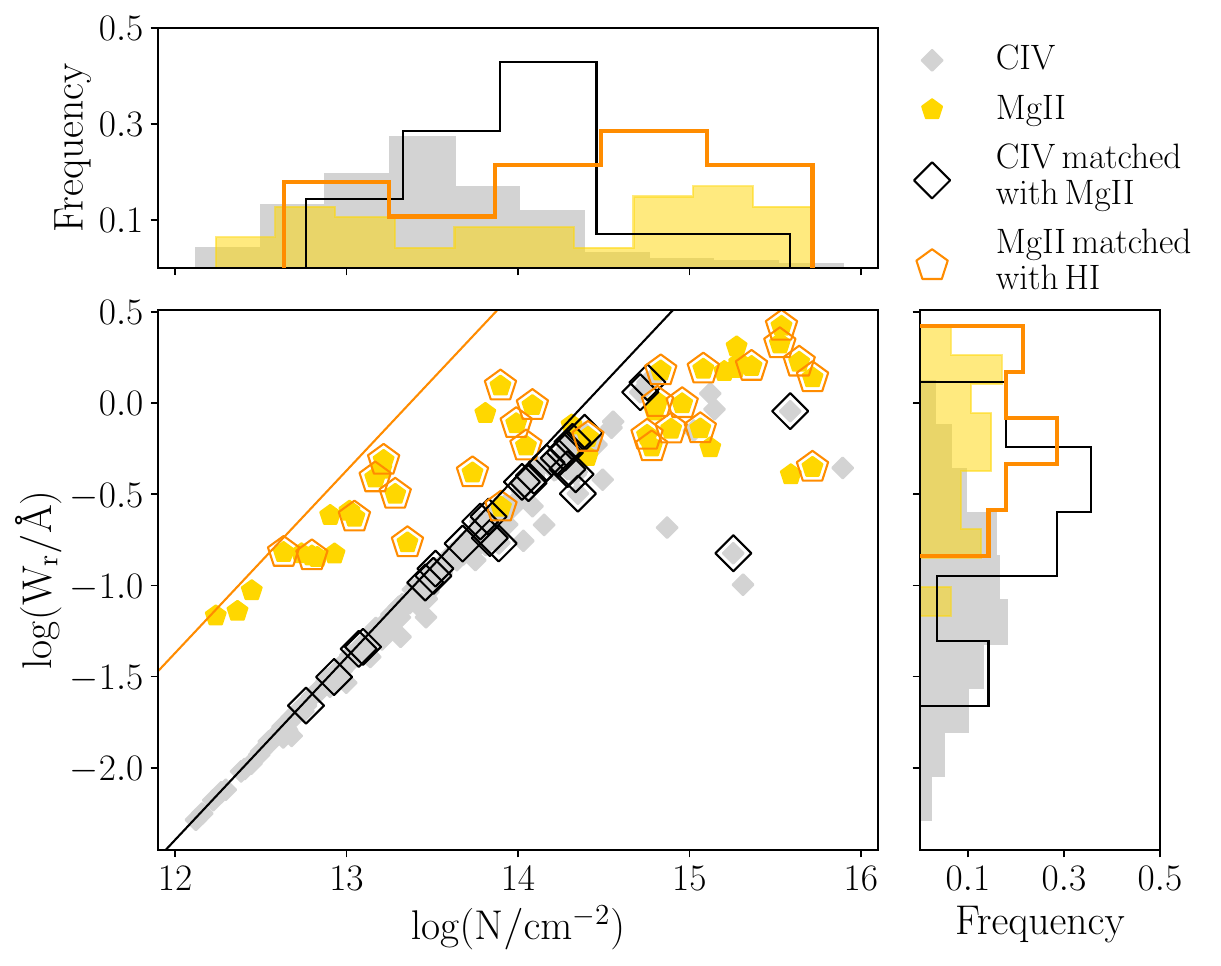}
\caption{Column density as a function of the rest-frame equivalent width of the absorbers for the full sample of \mgii\ (gold) compared to those that overlap with the strong HI systems from \maggiv\ (orange) and for the full sample of \civ\ lines from \maggv\ (gray) compared to those matched to the \mgii\ sample (black). Solid lines show the equivalent width as a function of the column density in the linear regime of the curve of growth. The distributions for column density and the equivalent width are shown on the right and top panels. We matched the samples within a linking velocity window of $|\Delta v|=150\rm\,km\,s^{-1}$.}
\label{fig:EWoverlap}
\end{figure}

The selection of \mgii, \civ\, and \hi\ absorbers in MAGG has been conducted independently, but these samples partially overlap. Knowing the degree of overlap between various ions helps compare this analysis with the results of previous series papers. 
To quantify this effect, we matched the \mgii\ and \civ\ samples in redshift space within $|\Delta v|=100,\,150,\,200,\,250\rm\,km\,s^{-1}$. The fraction of \civ\ absorbers matched to \mgii\ systems rises from $\approx12$ percent to $\approx14$ percent for increasing linking velocity, corresponding to $58-66$ percent of the \mgii\ lines sample ($28-31$ out of 47). In addition, independent of the linking velocity, $\gtrsim90$ percent of the \civ\ systems matched to the \mgii\ absorbers show larger rest-frame equivalent width (black datapoints and histograms in Figure \ref{fig:EWoverlap}) and width in velocity space compared the median values of the full \civ\ sample. Consistently, \citet{Steidel1992} and \citet{Schroetter2021} found that the strength of \civ\ systems that are not matched to any \mgii\ absorber is peaked towards lower equivalent widths due to the ionized gas being more diffuse than a neutral and metal-enriched phase around galaxies. 

We also caution that due to the difference in rest-frame wavelength, oscillator strength, and relative abundances, the \mgii\ absorption saturates at column densities $\approx1\rm\,dex$ smaller compared to the \civ. At fixed rest-frame equivalent width, the two ions do not trace the same gas density. In addition, we used {\sc cloudy} grids from \citet{Fumagalli2016}, constructed by assuming photo-ionization equilibrium, to verify that at typical CGM and IGM densities ($\log[(n_{\rm H}/{\rm cm^{-3}})]\lesssim-2.5$) \mgii\ absorbers are stronger than the \civ\ systems, with a rest-frame equivalent width $W_{\rm\mgii}\approx2\times W_{\rm\civ}$ in the linear regime. These expectations are consistent with the equivalent width ratios we directly measure from the matched \mgii\ and \civ\ systems exhibiting a similar number of components, as detailed in Appendix \ref{app:kinematics}, resulting in $W_{\rm\mgii}=(3.33\pm1.82)\times W_{\rm\civ}$. 
In this appendix, motivated by the detailed comparison of multiple gas phases at $z\approx2$ from \citet{Rudie2019}, we also explore the differences in the kinematics between low and medium ions that are aligned in redshift space and likely arise from the same gas clouds.

Due to the severe contamination of the sky emission in the NIR and the lower spectral resolution compared to the optical quasar spectra in which we have identified the \civ\ doublets, the sample of the \mgii\ absorbers includes only stronger systems. To quantify the effect of the different selection functions between the two samples, we estimated the $3\sigma$ upper limits on the rest-frame equivalent width of the ions at the redshift of LAEs that are not connected to any absorber. Taking the median value and the spread given by the 16th and 84th percentiles of the upper limits distribution, we obtain $W_{\rm r}^{\rm 3\sigma}=0.02\pm0.01\,$\AA\ for the \civ\ systems, $\approx3$ times lower and with a much smaller spread compared to the \mgii\ gas for which $W_{\rm r}^{\rm 3\sigma}=0.06\pm0.05\,$\AA. 

Finally, to connect the cool enriched gas phase traced by the \mgii\ absorbers to the neutral and dense gas, we consider the sample of 61 strong ($N_{\rm \hi}\gtrsim10^{16.5}\rm\,cm^{-2}$) hydrogen absorbers identified in MAGG by \cite{Lofthouse2023}. As done for \civ, we investigate the degree of overlap between the samples of \mgii\ and \hi\ absorbers. Depending on the choice of a linking velocity window ranging from $100\rm\,km\,s^{-1}$ to $250\rm\,km\,s^{-1}$, we report that the $57-60$ percent (corresponding to 27/47 to 28/47) of the \mgii\ systems overlap with a strong \hi. About 65 percent of these \mgii\ absorbers show a rest-frame equivalent width above the median of the entire sample (orange data points and histograms in Figure \ref{fig:EWoverlap}). At the same time, we do not observe any significant difference in the kinematics as traced by the profile width.


\section{Results}

\subsection{Searching for clustering of LAEs around \mgii\ absorbers}
\label{sec:MgII-LAE}

\begin{figure} 
\centering
\includegraphics[width=\columnwidth]{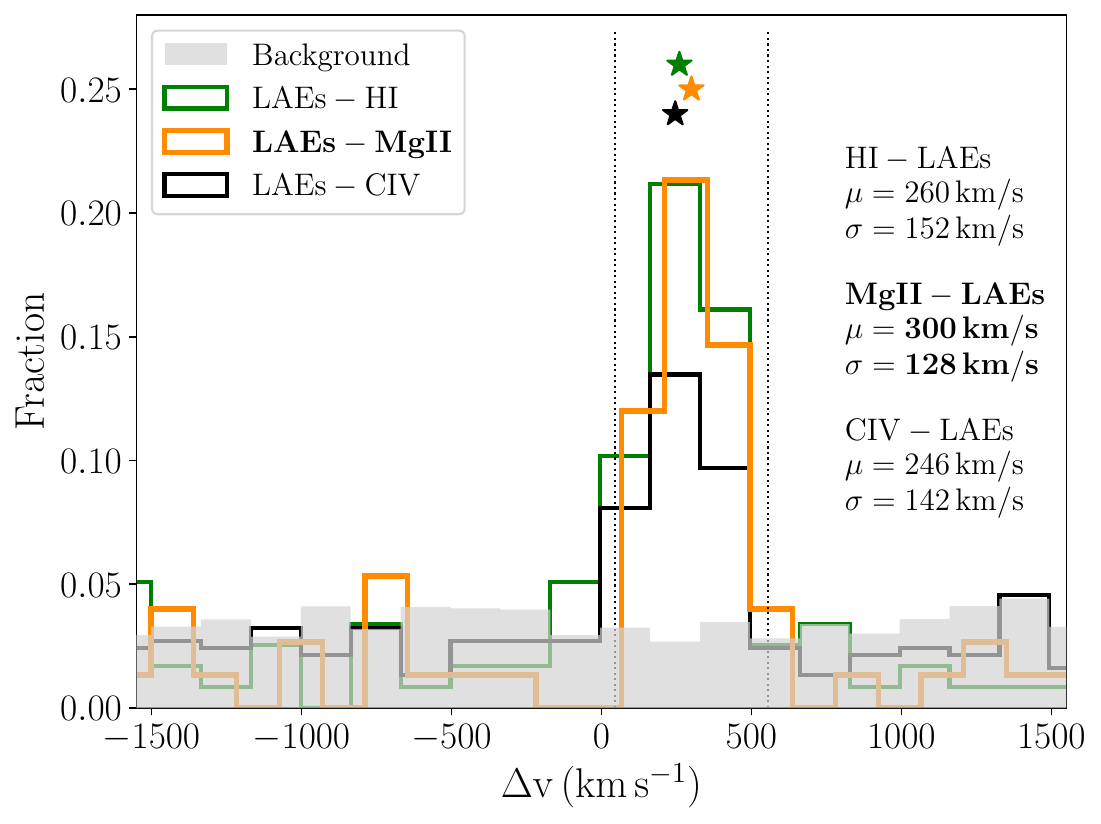}
\caption{Distribution of LAEs within $|\Delta v|\leq1500\rm\,km\,s^{-1}$ of \mgii\ systems (orange) as a function of the line-of-sight separation. As a comparison, we also show the LAEs connected to \civ\ absorbers from \maggv\ (black), those associated to strong \hi\ absorbers from \maggiv\ (green) and the distribution of a random sample representative of our expectations for the field (gray). The annotations show the best fit of the distributions considering a Gaussian function plus a background term, with the peaks shown as stars. The vertical dotted lines mark the $2\sigma$ interval around the peak, $\mu$, of the \mgii-LAEs distribution, where the largest excess of galaxies is found.}
\label{fig:dvR}
\end{figure}

We start our investigation into the physical connection between LAEs and the cool gas traced by the \mgii\ systems by cross-matching the two catalogs with velocity window $|\Delta v|\leq1500\rm\,km\,s^{-1}$. Figure \ref{fig:dvR} shows the galaxies' line-of-sight separation (left-hand panel) relative to the redshfit of the \mgii\ absorbers (orange). A clear peak is visible in the velocity distribution, suggesting LAEs are locally clustered around the \mgii. To gauge the significance of this detection, we reproduce the expectations for random regions in the Universe that are not selected via \mgii\ absorption (light gray). As described in \maggv, we achieve this by connecting LAEs observed in a random field to the \mgii\ absorbers found in a sightline randomly chosen between the remaining ones. We repeated this bootstrap procedure $10^3$ times. We then model the peak observed for the LAEs-\mgii\ connections with a Gaussian function plus an additional term to account for the background. We list the best-fit parameters in Figure \ref{fig:dvR}. 

According to these results, the majority of the LAEs are found to be clustered to the \mgii\ absorbers around $\mu\approx300\rm\,km\,s^{-1}$ and within $2\sigma\approx\pm256\rm\,km\,s^{-1}$ (vertical dotted lines in Figure \ref{fig:dvR}), while at larger velocities the distribution becomes consistent with the expectation for the random background. This also suggests that the velocity distribution is not symmetric around the redshift of the \mgii\ absorbers but significantly offset due to the systematic shift of the LAEs' redshift as a consequence of the resonant scattering of the Ly$\alpha$ photons. We thus correct the redshifts for the mean offset $z^\prime_{\rm Ly\alpha}=z_{\rm Ly\alpha}-\mu(1+z_{\rm Ly\alpha})^{-1}c^{-1}$. 

Having corrected the LAE redshifts for the velocity offset, we consider LAEs within $\pm500\rm\,km\,s^{-1}$ relative to the \mgii\ absorbers as associated with the cool gas and included in the following analysis. With this criterion, we identify 43 LAEs connected to 18 \mgii\ absorbers, corresponding to a detection rate of $\approx36\pm10$ percent (17/47) at $1\sigma$ confidence level. 
We also derive a first-order estimate of the amplitude of the clustering signal above the background by integrating the two distributions within the linking velocity window, finding a factor of  $\approx4$ of LAEs near \mgii\ absorbers compared to random expectation. We reserve a more statistical derivation of the overdensities of LAEs around the \mgii\ absorbers to the next section.  

We compare these results with the redshift space connection between LAEs and \civ\ systems from \maggv\ (black). In \maggv\, we reported a detection rate of $36\pm5$ for the $z>3$ \civ\ absorbers, consistent within $1\sigma$ with the result we obtained for the \mgii\ gas. The excess of LAEs within $\pm500\rm\,km\,s^{-1}$ from the \mgii\ absorbers is a factor $\approx\times1.5$ larger compared to the \civ-LAE clustering amplitude, similar to the clustering of LAEs around strong \hi\ absorbers from \maggiv\ (green in Figure \ref{fig:dvR}). Given the results from Section~\ref{sec:overlap_MAGG}, we understand this finding because the \mgii\ sample significantly overlaps with optically thick absorbers that trace higher overdensities more correlated to galaxies. On the other hand, the overlap between \mgii\ and \civ\ absorbers is only partial and skewed to higher equivalent widths. The weak \civ\ absorbers that, based on \maggv\ results, trace a diffuse enriched medium that is not always connected to galaxies dilute the clustering signal. 

\subsubsection{A lower LAE detection rate for \mgii\ gas}

\begin{figure} 
\centering
\includegraphics[width=\columnwidth]{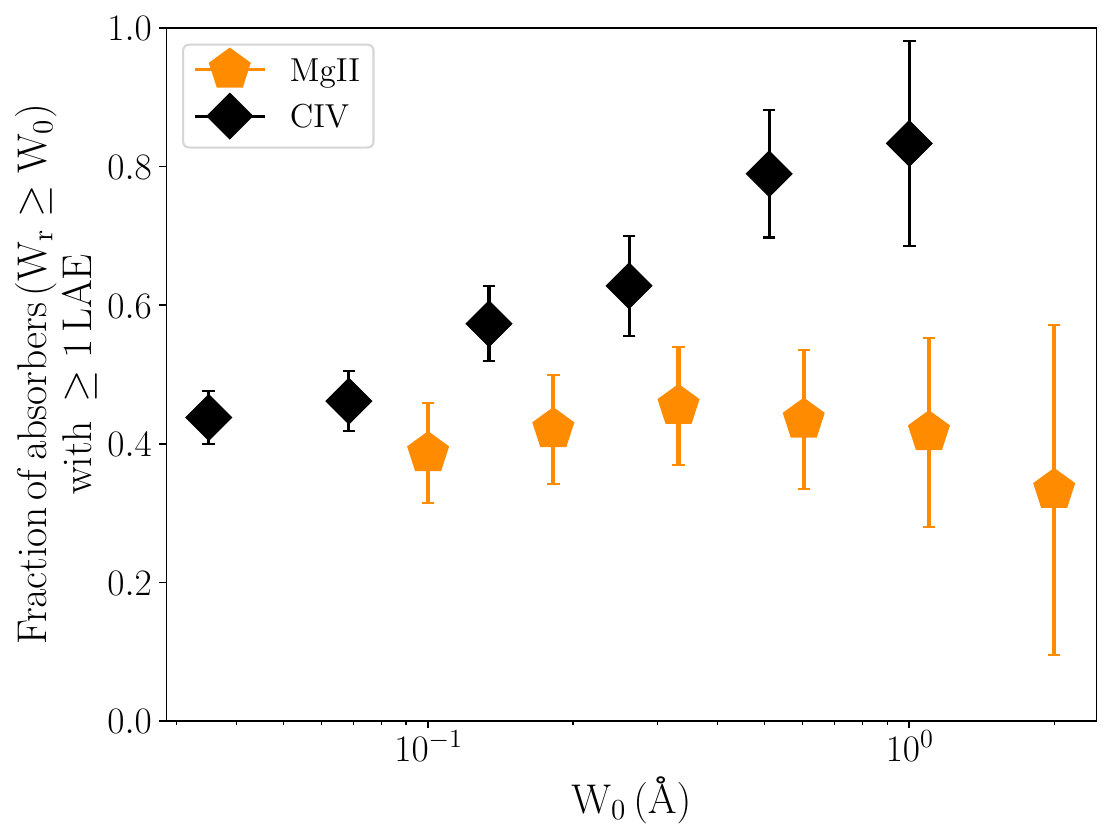}
\caption{Fraction of \mgii\ (orange) and \civ\ (black) absorbers with equivalent width $W_{\rm r}\geq W_0$ that are found to be connected to at least one LAE within $\pm500\rm\,km\,s^{-1}$ as a function of the threshold $W_0$.}
\label{fig:det_rate}
\end{figure}

The rate at which we can find one or more galaxies close in velocity to the absorbers provides information about which type of galaxies are surrounded by different gas phases. We thus estimate the fraction of absorbers that are connected to at least one LAE within $\pm500\rm\,km\,s^{-1}$, as a function of their rest-frame equivalent width, $W_{\rm r}\geq W_0$. As shown in Figure \ref{fig:det_rate}, the detection rate increases monotonically with the strength of the \civ\ absorbers (black). Instead, we do not find any significant trend for the \mgii\ gas (orange). To strengthen this analysis with a more statistical evidence, we perform the Spearman-rank correlation test. It returns the degree of monotonic correlation, expressed by the correlation coefficent $\rho$, and the probability that no monotonic correlation exists between the variables, $p-$value\footnote{The closer the correlation coefficient $\rho$ is to the values $\pm1$, the stronger the monotic correlation. $|\rho|<0.20$ is considered a very weak correlation, while a correlation is strong for $|\rho|>0.60$. On the other hand, the significance of the correlation depends on the $p-$value: if $p-$value$<0.05$, the correlation is significant at least at $2\sigma$ level.}. The Spearman-rank correlation test results in $\rho=0.99$, $p-$value$=0.003$ for the \civ\ and $\rho=-0.26$, $p-$value$=0.63$ for the \mgii\ gas, supporting these conclusions. 

To further explore this difference, we focus on the strong $W_{\rm r}^{\rm\mgii}\geq1\,\text{\AA}$ \mgii\ absorbers that have corresponding kinematically-aligned \civ\ systems (see the Appendix \ref{app:kinematics}). In this subsample, we observe that systems linked to at least one LAE have 
on average strong \civ\ equivalent width. Conversely, \civ\ absorbers with $W_{\rm r}^{\rm\civ}\lesssim0.6\text\,{\AA}$ are less correlated to galaxies. In this case, the weakness of the \civ\ absorbers and the strength of the \mgii\ systems suggest either an excess of cool gas or a lower ionization; indeed, these systems largely overlap with the DLA population in MAGG.

This lower detection rate in strong \mgii\ absorbers is unexpected, as this gas is largely believed to trace overdense and neutral regions in close proximity to galaxies. We thus explore the possibility that the galaxies connected to these strong absorbers lie at close separation from the quasar sightline \citep[see, e.g.,][]{Fynbo2010, Fynbo2011, Nielsen2013b_general, Schroetter2019, Guha2023}. We project 14 layers around the wavelength of the Ly$\alpha$ emission at the redshift of the absorbers (corresponding to about $\pm1000\rm\,km\,s^{-1}$, on average) to build pseudo-narrow band cut-outs around the central quasar of each field from the quasar point spread function (PSF) and continuum-subtracted datacubes. By visually inspecting these images, we did not find any Ly$\alpha$ emitting galaxies that were not identified by our selection due to their proximity or overlap with the quasar PSF up to impact parameters of $20\rm\,kpc$. 

Having excluded the presence of UV bright or \lya\ emitting galaxies, a plausible explanation is the existence of a dusty galaxy population excluded from our UV selection. At present, there are no systematic investigations into a high-redshift obscured population near strong \mgii\ absorbers, an area of parameter space that should be explored with the Atacama Large Millimeter/submillimeter Array (ALMA) and the {\it James Webb Space Telescope} (JWST). At lower redshift, \citet{Menard2012} studied the dust content of strong $W_{\rm r}\geq0.8\,\text{\AA}$ \mgii\ absorbers. At $z\approx0.5$, they found that it is similar to the dust content of galactic disks and showed that the \mgii\ absorbers carry $\approx1/2-1/3$ of the total dust expelled by galaxies. Similarly, \citet{Budzynski2011} reported that in SDSS the $\approx20$ percent of $W_{\rm r}\geq1\,\text{\AA}$ \mgii\ systems are missed due to heavy dust obscuration. Once selected by high-column density hydrogen, \citet{Neeleman2017, Neeleman2018, Neeleman2019} reported a high detection rate of [\ion{C}{ii}] emitters around clumpy, neutral and enriched regions at $z\approx4$, attributing the previous non-detection of galaxies to the significant dust obscuration.

Systematic follow-up of a sample of strong \mgii\ absorbers with high resolution and at IR/submm frequencies will therefore be instrumental to test the presence of UV-dark galaxies, or galaxies lying closer than $\approx0.8\rm\,arcsec$ to the sightline.

\subsection{Characterising the LAE overdensity around \mgii\ absorbers}
\label{sec:connection_with_LAEs}

\begin{figure*} 
\centering
\includegraphics[width=\columnwidth]{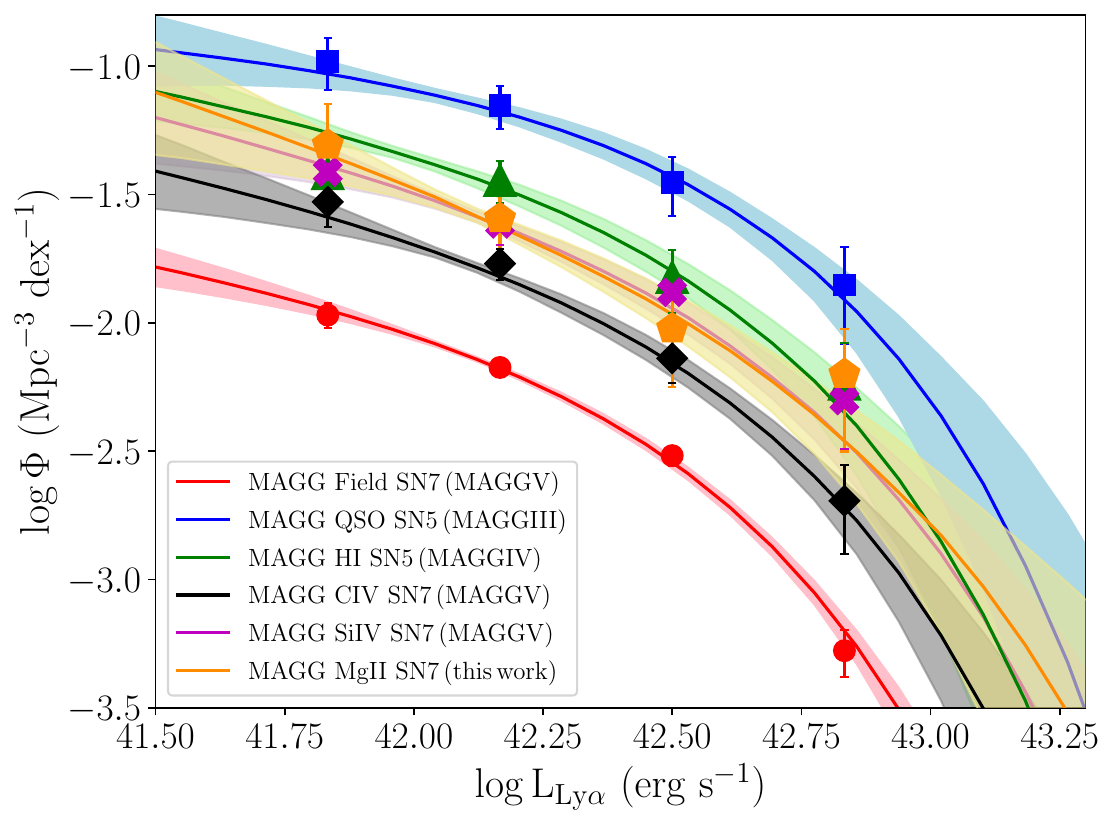}
\includegraphics[width=\columnwidth]{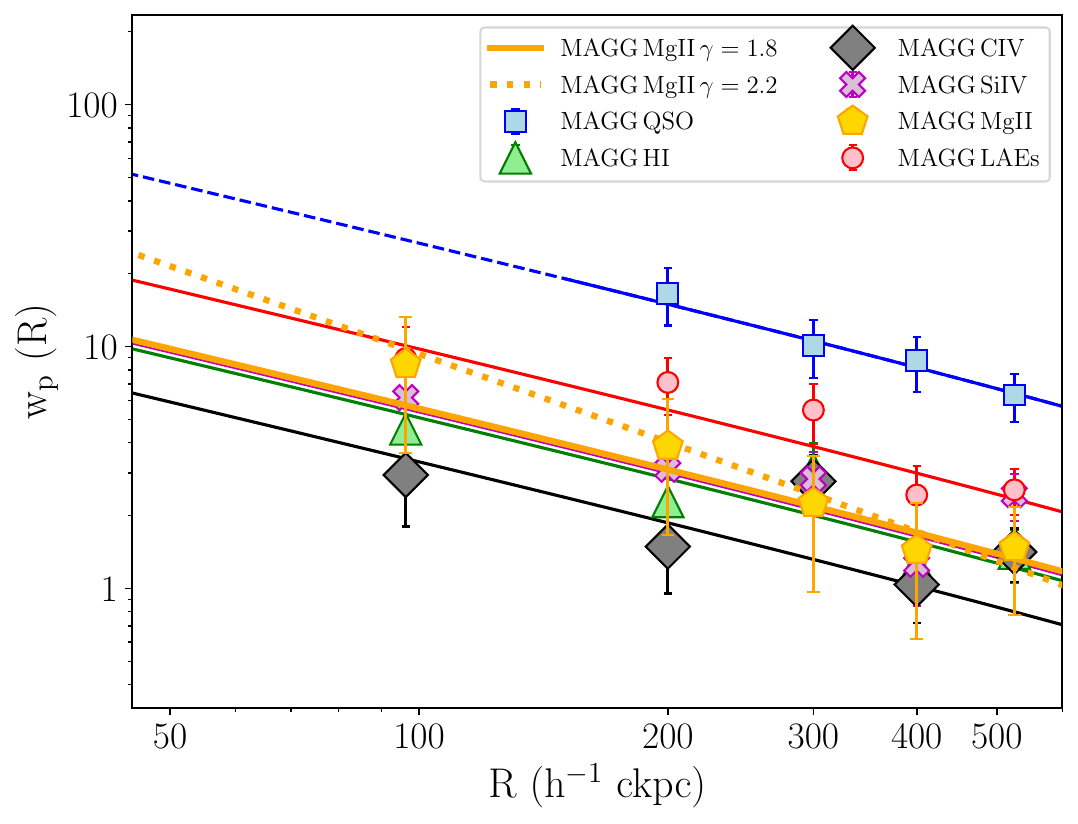}
\caption{Left panel: LAE luminosity function within $|\Delta v|\leq500\rm\,km\,s^{-1}$ of the \mgii\ absorbers (orange) compared to the field control sample (red, \maggv) and to the results for LAEs around other tracers studied in MAGG (high-column density hydrogen absorbers in green from \maggiv, \civ\ and \siv\ systems in black and purple from \maggv\ and quasars in blue from \citealp{Fossati2021}). We show the parametric (line) and non-parametric (data points) estimates for each sample. Right panel: Projected cross-correlation function for LAEs within $|\Delta v|\leq500\rm\,km\,s^{-1}$ of the \mgii\ absorbers (orange) assuming $\gamma=1.8$ (solid line) and $\gamma=2.2$ (dotted line). The other populations explored in MAGG are shown with the same colors adopted in the left panel, while the LAEs auto-correlation function derived in \maggv\ is shown in red. Vertical error bars are derived by bootstrapping the redshift of the \mgii\ absorbers over 1000 iterations.}
\label{fig:LF_Xcorr}
\end{figure*}

The peak in the line-of-sight separation between LAEs and \mgii\ absorbers suggests that the low-mass galaxies are preferentially clustered around regions harboring the cool gas rather than inhabiting random regions of the Universe. This aligns with results from earlier MAGG papers (IV and V), which consistently point to a preference for LAEs in tracing overdense cosmic structures selected by strong ions in quasar spectra. Following the methods of previous MAGG analyses, we quantify this link with the luminosity function of LAEs connected to \mgii\ absorbers at $z\approx3$.
Referring to the formalism described by \citet{Fossati2021}, we compare the non-parametric estimator $1/V_{\rm max}$ \citep{Schmidt1968, Felten1976} to a parametric Schechter \citep{Schechter1976} fit of the unbinned data. Incompleteness is accounted for by 
performing injection-recovering simulations of 500 mock LAEs for 1000 iterations over all the MAGG fields.
The lensed field J142438+225600 \citep{Patnaik1992} is excluded from the analysis. As a final step, we compute the effective volume by considering only the velocity window we inspected to connect the absorbers to the galaxy. Once the LAE redshifts are corrected for the $\mu\approx300\rm\,km\,s^{-1}$ offset due to the radiative transfer effects, this corresponds to $\Delta v=\pm500\rm\,km\,s^{-1}$ around the redshift of each \mgii\ absorbers. We then bin all the galaxies with a step of $0.2\rm\,dex$ in luminosity, assuming that the luminosity function of LAEs does not depend on the redshift \citep{Herenz2019}. The non-parametric $1/V_{\rm max}$ estimator and the parametric fit are both shown in Figure \ref{fig:LF_Xcorr} (orange points and solid line, respectively). For the luminosity function of LAEs around the \mgii\ absorbers we obtain the following best-fit parameters for a Schechter \citep{Schechter1976} function 
\begin{equation}
\begin{split}
    \log[\phi^\star/({\rm Mpc^{-3}dex^{-1}})]=-2.307\pm0.760 \\
    \log[{L}^\star/({\rm erg\,s^{-1}})]=42.811\pm0.497 \\
    \alpha=-1.665\pm0.460    \:.
\end{split}
\end{equation}


To quantify the amplitude of the clustering of LAEs within $\pm500\rm\,km\,s^{-1}$ around the \mgii\ absorbers, we integrate the luminosity function to compute the LAEs number density and compare it relative to the expectation for random regions of the Universe, represented by the 'MAGG field' sample that includes 617 galaxies at $3.0<z<4.5$, a range chosen to be consistent with the redshift distribution of the absorbers. In a similar way as done with the clustering in velocity space in Section \ref{sec:connection_with_LAEs}, we thus estimate that the surroundings of the \mgii\ absorbers are a factor of $\approx4$ overdense in LAEs compared to random regions of the Universe. 

Compared to different gas tracers (\civ\ and \hi\ absorbers from MAGG IV, V) and the quasar population from \citet{Fossati2021}, we observe no difference in the shape of the luminosity function, implying that the same galaxy population is identified around different tracers, with no excess in a given luminosity interval. Instead, we observe differences in normalization, as already noted in the previous MAGG papers. The LAEs near \mgii\ absorbers, in particular, have a $\approx\times1.3$ lower normalization compared to \hi\ systems. Therefore, the two populations largely overlap, which is unsurprising given that both ions trace a cool and partially neutral medium. LAEs linked to \mgii\ appear to have a marginally ($\approx\times1.7$) higher normalization compared to around the \civ\ gas, which - as done in MAGG V - we attribute to the dilution in the signal induced by lower equivalent width \civ\ absorbers being less correlated to galaxies.

To complete the picture of the overdensity of LAEs around the \mgii\ gas, we follow the method described by \citet{Trainor2012} and computed the reduced angular cross-correlation function. 
Based on the random sample described in Section \ref{sec:MgII-LAE}, we estimated the surface density of randomly distributed LAEs connected to \mgii\ absorbers within $|\Delta v|\leq500\rm\,km\,s^{-1}$ to be of the order of $\Sigma_{\rm rand}=(8.160\pm2.896)h^{2}\rm cMpc^{-2}$. We fit the data with the reduced angular cross-correlation function defined as:
\begin{equation}
    \omega_{\rm P}(R)=(R/r_{0})^{\gamma}\times\,_{2}F_{1}(1/2,\gamma/2;-3/2;-R_{\rm los,0}^{2}/R^{2})\:,
\label{eq:wp_model}    
\end{equation}
where $R$ is the projected separation between the LAEs and the central quasar of each field, $_{2}F_{1}$ is the Gaussian hypergeometric function and $R_{\rm los,0}$ is half the amplitude of the redshift window we searched for gas-galaxies connection, in physical units \citep[see][]{Fossati2021}.
We fixed the power law index $\gamma=1.8$, as done in similar studies (\citealp[see e.g.][]{Diener2017, GV2019, Fossati2021}), to reduce the degeneracy with normalization. We derive a clustering length $r_0=1.53^{+0.43}_{-0.48}h^{-1}\rm cMpc$, $\approx1.3$ times higher than the \civ\ gas from \maggv, but still consistent within $1\sigma$ uncertainties. The difference between the two gas phases is larger within $R\lesssim250\,h^{-1}\rm\,ckpc$ and decreases for increasing distance from the galaxies. A steeper power-law index, which we fix to $\gamma\approx 2.2$ in line with the best-fit model found by \citet{Dutta2021} for the \mgii\ absorbers at $z\lesssim2$, yields a lower correlation length, $r_0=1.32^{+0.29}_{-0.33}h^{-1}\rm\,cMpc$, and better capture the cross-correlation for small angular separations. On the other hand, values of $\gamma>1.8$ imprecisely model the flatter cross-correlation function of the \civ\ gas. As suggested by \citet{Dutta2021}, this is consistent with the strong \mgii\ absorbers to be preferentially concentrated close to the associated galaxies compared to the more ionized gas phases. Indeed, we show in Figure \ref{fig:Disc:EW_R} this seems to be the case also at higher redshift. In Figure \ref{fig:LF_Xcorr}, we also compare with the other gas tracers introduced in previous MAGG papers, for which fitting parameters can be found in table 7 of \maggv. Indeed, the clustering length of LAEs around the \mgii\ systems is consistent with those of LAEs clustered to LLSs and SiIV absorbers and $\approx\times1.4$ smaller than the self-clustering of LAEs, although still consistent within $1\sigma$ level due to the large uncertainties and the limited statistics of the samples.

Combining all these findings from the MAGG survey, we find that the number density of LAEs surrounding the \civ\ absorbers is lower relative to the \mgii\ and the LLSs at any projected separation. This is expected considering that the \hi\ and the \mgii\ absorbers trace a similar cool gas phase, although with different densities, and that the MAGG \mgii\ sample is selecting preferentially strong partially-neutral systems. In \maggv, we attributed the difference in the ionized phase to the weaker \civ\ absorbers, which appear less connected to galaxies and to the denser medium selected by lower-ionisation ions. These results suggest that the LAE number density around different tracers depends on the absorbers' strength and ionization, which vary with different gaseous components, such as the galaxies' inner CGM, the overdense gas along the filaments, or an enriched and diffuse medium at larger distances. Moreover, different gas tracers are not fully independent: within this picture, LAEs - by virtue of their high number density - become themselves tracers of cosmic structures on a range of scales, containing multiple galaxy populations and gas phases.

\subsection{The effect of the large-scale galaxy environment}
\label{sec:effect_of_groups}


So far, we have established and explored the physical connection between LAEs and different gas phases traced by \mgii\ and \civ\ absorbers. The IFU observations included in the MAGG survey identified more than 150 groups of LAEs across the 28 quasar fields (see Section \ref{sec:catalogs_MAGG}). Throughout this work, we recognize as part of a group LAEs with at least one other galaxy within $\pm500\rm\,km\,s^{-1}$ in redshift space. This makes it possible to introduce an additional dimension in studying the gaseous surroundings of LAEs: the large-scale galaxy environment. In previous works, it has been shown that both the strong \hi\ \citep{Muzahid2021, Lofthouse2023} and the \civ\ absorption \citep{Muzahid2021, MAGGV, Banerjee2023} is enhanced in groups compared to around isolated LAEs at $z\approx3-4$. At lower redshift $z\approx0.5-2.0$ a similar effect has been observed in MUSE surveys \citep[e.g.][]{Peroux2019, Fossati2019, Dutta2020, Dutta2021}. 

With our new \mgii\ sample, we complete this picture by examining the role played by the large-scale galaxy environments on the cool gas and compare it to the effect on the more ionized phase traced by the \civ\ systems. Out of all the \mgii\ absorbers associated to LAEs within a line-of-sight separation $|\Delta v|\leq500\rm\,km\,s^{-1}$, we find that the $65\pm11$ percent is connected to galaxies that are part of a group, while only the remaining fraction is associated to isolated ones. In this section, we explore the impact of the large-scale environment on different gaseous phases as a function of the galaxies' properties and the projected separation.

\subsubsection{Covering fraction of different gas phases}
\label{sec:env_CF}

\begin{figure*}
\centering
\includegraphics[width=\textwidth]{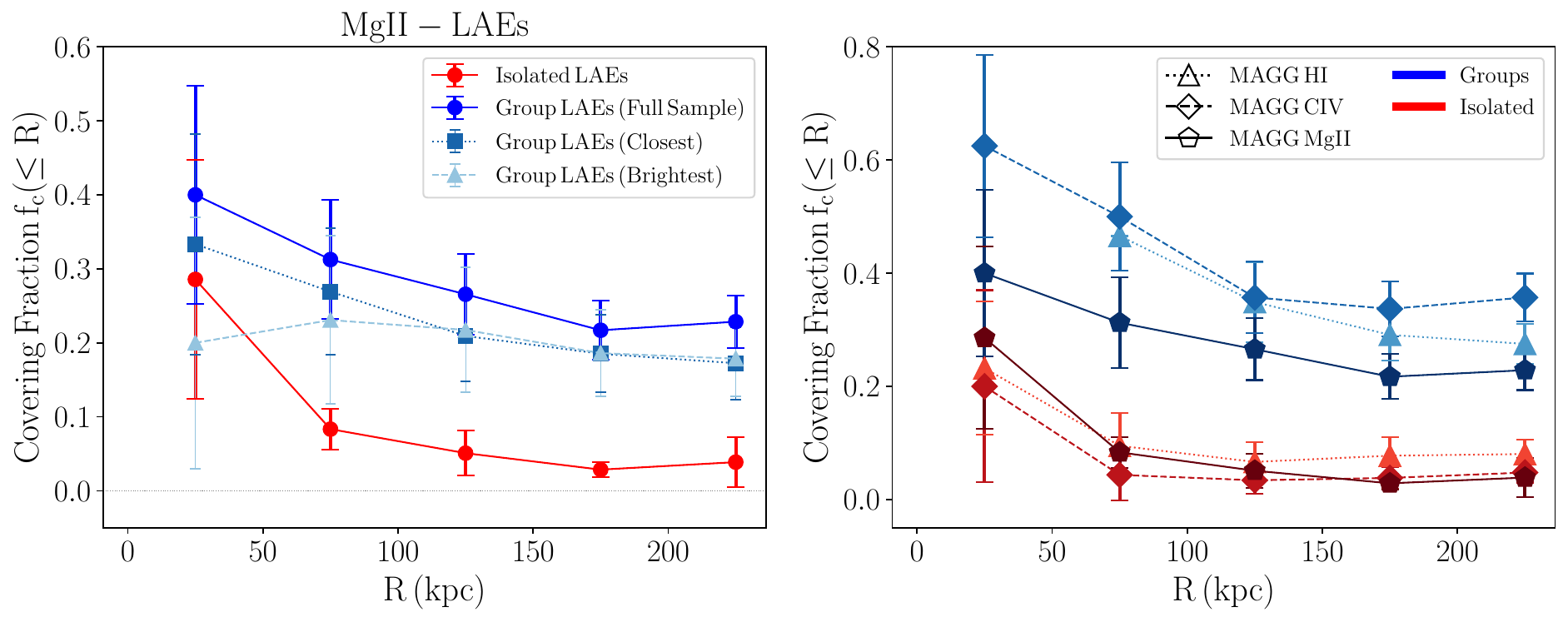}
\caption{Left panel: Cumulative covering fraction of LAEs connected to $W_{\rm r}\geq0.10\,$\AA\ \mgii\ absorbers as a function of the projected separation. We distinguish between isolated (red) and group (shades of blue) galaxies. The results obtained considering only the brightest (triangles) and the closest (squares) member of each group are consistent with each other and with the full sample within $1\sigma$. Right panel: Comparison between the cumulative covering fraction for isolated (red) and group (shades of blue) LAE around the strong \hi\ hydrogen absorbers (triangles), the \civ\ (diamonds) and \mgii\ (pentagons) systems, assuming $N_{\rm\hi}>10^{17.5}\rm\,cm^{-2}$ and $W_{\rm r}\geq0.30\,$\AA.}
\label{fig:CFgroups}
\end{figure*}

We compute the cumulative covering fraction, $f_{\rm c}(\leq R)$, which is defined as the probability of detecting absorbers with $W_{\rm r}\geq W_0$ within a projected separation $R$ and line-of-sight velocity in the range $|\Delta v|\leq500\rm\,km\,s^{-1}$ around a galaxy. We split the sample of galaxies connected to the \mgii\ absorbers into groups (34 LAEs that are part of 11 different groups) and isolated (6 LAEs). In computing the covering fraction, we only consider LAEs connected to \mgii\ absorbers with a rest-frame equivalent width above the threshold $W_0=0.10\,$\AA, i.e., the completeness limit. The $1\sigma$ uncertainties account for the 68 percent Wilson-score confidence intervals. 
In \maggiv, as sightlines were selected to host \hi\ absorbers, we accounted for this selection bias by renormalizing the number of systems per unit redshift by the corresponding cosmological value, $\ell(X)=dN/dX$. For metal absorbers, due to the higher number of systems per sightline and the fact that we do not explicitly select these populations, we find that the incidence of \mgii\ and \civ\ absorbers in MAGG is comparable to the cosmic values within uncertainties, and hence we do not apply additional renormalizations.

Results in Figure \ref{fig:CFgroups} (left panel) show that the cool gas traced by the \mgii\ absorption is $\approx\times4$ more likely to be found around galaxies in groups than around isolated LAEs. At $R\leq100\rm\,kpc$ the covering fraction is $f_{\rm c}=0.27\pm 0.09$ around group LAEs considering only the closest member, and  $f_{\rm c}=0.08\pm 0.03$ around isolated LAEs, decreasing to $f_{\rm c}=0.17\pm 0.05$ and $f_{\rm c}=0.04\pm 0.03$ at $R\leq250\rm\,kpc$ for the two subsamples, respectively. Taking either the full sample of groups or the brightest component leads to results that are all consistent within $1\sigma$ (see the left panel in \ref{fig:CFgroups}). The observed excess of \mgii\ absorption in groups is consistent with the findings at $z\lesssim2$ by \citet{Nielsen2018_grp} and \citet{Dutta2020, Dutta2021}. Section~\ref{sec:disc_zevolution_grp} discusses more details on the redshift evolution.

At $z\lesssim2$, \citet{Dutta2021} has shown that the galaxy environment impacts the gas distribution differently according to the gas phase. We thus compare explicitly the cool and enriched gas phase traced by the \mgii-bearing gas, the neutral phase traced by strong \hi\ systems, and the more ionized gas traced by \civ (right panel in Figure \ref{fig:CFgroups}). For this analysis, we set the equivalent width threshold to $W_0=0.3\,\text{\AA}$ to limit the effects of the different selection functions of the \mgii\ and \civ\ samples. This corresponds to a difference in column density of $\log(N_{\rm\civ}/{\rm cm^{-2}})-\log(N_{\rm\mgii}/{\rm cm^{-2}})\approx1\rm\,dex$ in linear regime. The probability of observing any of these absorption systems is $\approx3-4$ times higher in groups compared to isolated LAEs, regardless the choice of the sub-sample of groups. In detail, if, on the one hand, the \civ\ and \hi\ gas show a similar covering fraction around group LAEs, on the other hand, that of the \mgii\ absorbers is consistent with the other tracers within $1\sigma$ confidence level, but $\approx25$ percent lower despite the large degree of overlap among the different samples in MAGG. This may depend on the equivalent-width threshold, which selects relatively strong \civ\ (corresponding to the 84th percentile of the \civ\ equivalent width distribution), and intermediate equivalent-width \mgii\ absorbers (25th percentile). However, at $z\lesssim2$ \citet{Dutta2021} found that the \mgii\ covering fraction is $\approx2$ times lower and steeper compared to that of the \civ\ gas, which flattens at large impact parameters. They interpreted these differences as due to the warm and ionized \civ\ gas being more extended around galaxies than the \mgii\ cold phase. Altogether, these results support a picture in which the effect of the large-scale galaxy environments on the surrounding gas depends on the gas phase. 

Motivated by the previous studies at lower redshift and by our findings in \maggv\ for LAEs around the \civ\ absorbers, we compare the properties of groups to those of the isolated LAEs to verify if any intrinsic difference in the Ly$\alpha$ luminosity or projected separation of these two galaxy populations may be responsible for enhancing the metal absorption in LAEs overdensity. To this end, we performed a KS-test, finding a $p$-value$\gtrsim0.40$ and $p-$value$\gtrsim0.13$ ($p$-value$\gtrsim0.54$ and $p-$value$\gtrsim0.36$), depending on the different sub-sample of groups, when comparing the impact parameter and the \lya\ luminosity of isolated galaxies and of those in groups that are connected to \mgii\ (\civ) absorbers, respectively. These values suggest no statistical difference between the distributions of luminosity and impact parameters of groups and isolated LAEs. This result clears the way for two possible scenarios to explain the excess of absorption in groups: the cooler \mgii\ gas may be either affected by the large-scale structure in which groups reside, as found for the ionized \civ\ phase at $z\gtrsim3$ in \maggv\ \citep[see also][]{Muzahid2021}, or is more sensitive to the local galaxy environment and shaped by the gravitational and hydrodynamic interactions that occur in galaxy overdensities, as found for the \mgii-bearing gas at $z\lesssim2$ \citep[see, e.g.,][]{Dutta2021}.

\subsubsection{Environmental effects on the absorption properties}
\label{sec:env_abs_prop}

\begin{figure} 
\centering
\includegraphics[width=\columnwidth]{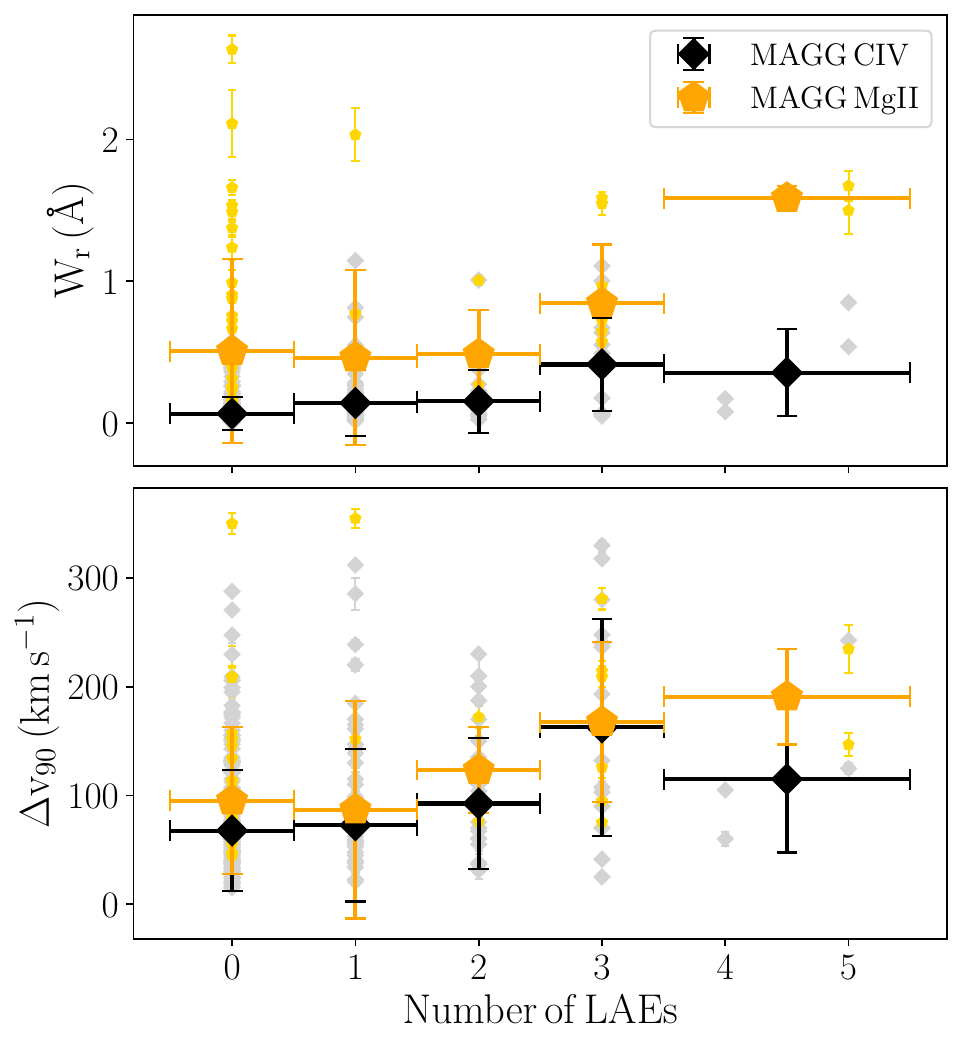}
\caption{Effect of the large-scale galaxy environment on the absorption properties. We show the absorbers' rest-frame equivalent width (upper panel) and the width in velocity space (lower panel) as a function of the number of LAEs connected to the absorbers. Individual \civ-LAE (gray) and \mgii-LAE (gold) connections are shown, as well as the binned median and standard deviation (black and orange, respectively).}
\label{fig:Prop_Ngal}
\end{figure}

Having observed that the galaxy environment affects the covering factor near LAEs, we now investigate whether the number of LAEs connected to each individual \mgii\ and \civ\ system plays any role in shaping the strength of the absorption and the kinematics of different gas phases. In Figure \ref{fig:Prop_Ngal}, we show the rest-frame equivalent width (upper panel) and $\Delta v_{90}$ (lower panel) as a function of the number of LAEs within $\pm500\rm\,km\,s^{-1}$ of the absorbers (grey points). The medians (data points) and standard deviations (error bars) are also shown. In \maggv, we showed how \civ\ systems connected to $\geq2$ LAEs are, on average, stronger and mildly more kinematically complex than those associated with $\leq1$ galaxy. Taking into consideration that the sample of \mgii\ lines tends to select stronger systems, we observe a trend where both the strength and the kinematics of the \mgii\ absorbers rise with the number of connected LAEs, exceeding those of the \civ\ systems in the richest groups. 

To support this evidence, we performed a non-parametric Spearman rank correlation test to quantify the degree of monotonic correlation between the properties of the absorbers and the number of associated LAEs, together with the probability of no correlation. For the \mgii\ absorbers, we measure correlation coefficients of $\rho\approx0.30$ and $p-$value$\approx0.045$ and $\rho\approx0.27$ and $p-$value$\approx0.049$ for the equivalent-width and $\Delta v_{90}$. For the \civ\ systems, we find $\rho\approx0.39$ and $p-$value$\approx0.001$ and $\rho\approx0.22$ and $p-$value$\approx0.002$ for the equivalent-width and $\Delta v_{90}$. The positive correlation of the strength and the kinematics of the \mgii\ and \civ\ lines with the number of LAEs is thus statistically significant for the two ions. At $z\approx0.5$, \citet{Lan2014} similarly found a positive correlation between the strength of the \mgii\ absorbers and the mean number of star forming-galaxies at small impact parameters. Likewise, at $z\lesssim1$, \citet{Dutta2021} reported that the strength and the incidence of the \mgii\ absorption increase with the galaxies number overdensity \citep[see also][]{Dutta2020} but found only a weaker dependence of the \civ\ covering fraction and equivalent width on the large-scale galaxy environment (see Section \ref{sec:disc_zevolution_grp} for a detailed discussion).

\begin{figure} 
\centering
\includegraphics[width=\columnwidth]{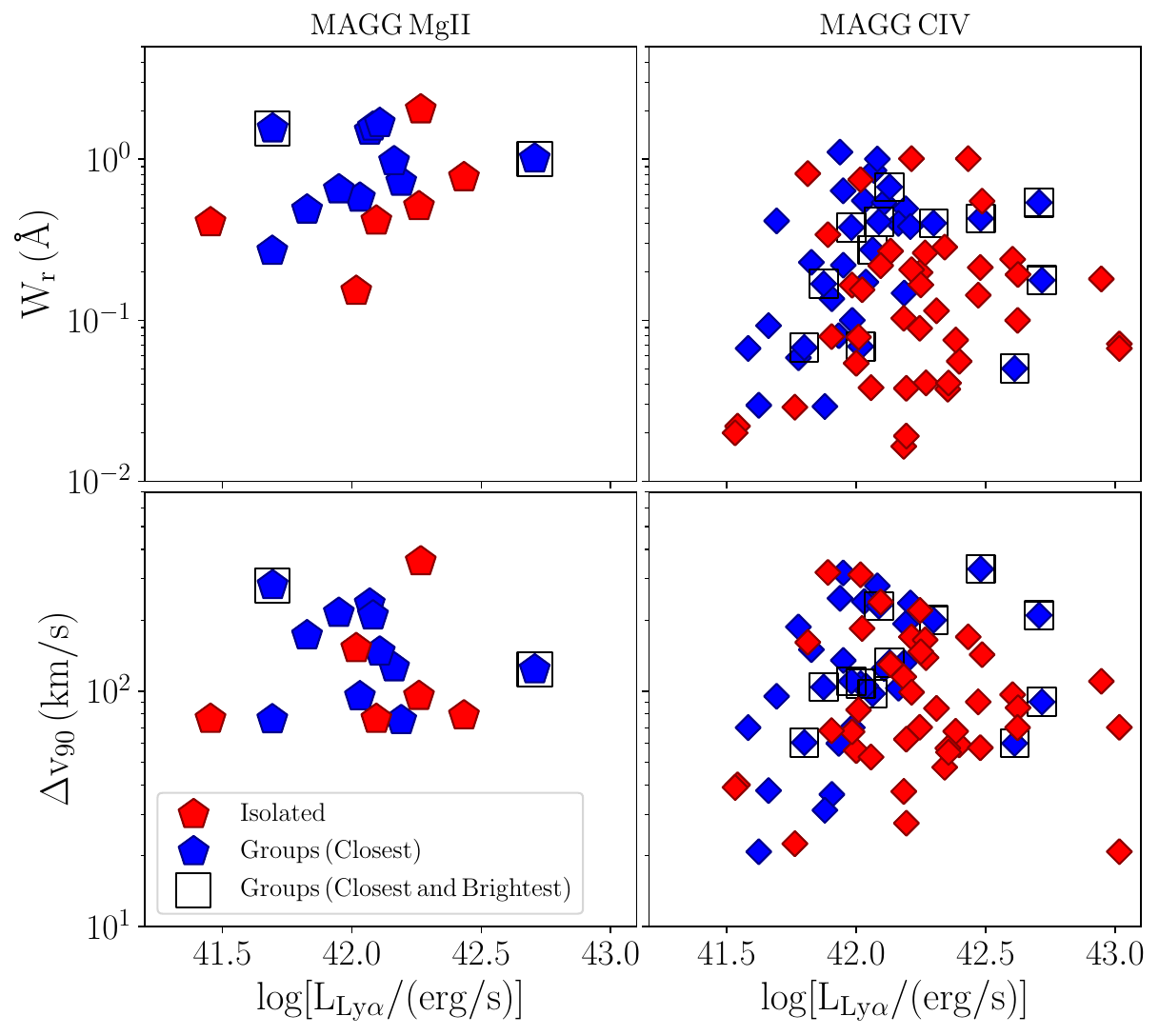}
\caption{\mgii\ (left panels) and \civ\ (right panels) absorbers' rest-frame equivalent width (top panels) and width in velocity space (bottom panels) as a function of the Ly$\alpha$ luminosity for isolated galaxies (red) and the closest member of each group (blue). Empty squares mark the case in which the closest galaxy of a group is also the brightest. The results from the Spearman rank test to assess the presence and the significance of possible correlations are listed in Table \ref{tab:GRPcorr}.}
\label{fig:Prop_GRP_ISO}
\end{figure}

Besides the overdensity of galaxies clustered to the absorbers, we also study whether the intrinsic property of Ly$\alpha$ luminosity shows trends with the absorption characteristics of different gas phases. We thus directly correlate, in Figure \ref{fig:Prop_GRP_ISO}, the strength of the absorption (upper panels) and the kinematics (bottom panels) with the luminosity of isolated galaxies (red) and groups (blue). Reported in Table \ref{tab:GRPcorr}, as a statistical diagnostic, are the results of the non-parametric Spearman's rank correlation test measuring the degree of monotonic correlation, $\rho$, and the probability that the two variables are not correlated, the $p-$value. We tested the correlation considering for each group the brightest galaxy and the one closest to the line-of-sight separately, as we noticed that they coincide only in 2/11 cases for LAEs around the \mgii\ and in 12/34 cases for those around the \civ\ absorbers).

\begin{table}
\centering
\begin{tabular}{cccc}
\hline
Line & LAEs & $W_{\rm r}\,$(\AA) & $\Delta v_{90}\,\rm(km\,s^{-1})$ \\
\hline
\mgii & Closest       & $\rho=0.43\,p=0.05$ & $\rho=0.34\,p=0.31$ \\
      & Brightest     & $\rho=0.54\,p=0.03$ & $\rho=0.04\,p=0.89$ \\
      & Full          & $\rho=0.06\,p=0.72$ & $\rho=0.47\,p=0.64$ \\
      & Isolated      & $\rho=0.89\,p=0.02$ & $\rho=0.31\,p=0.54$ \\
\hline
\civ  & Closest       & $\rho=0.43\,p=0.01$ & $\rho=0.42\,p=0.01$ \\
      & Brightest     & $\rho=0.33\,p=0.06$ & $\rho=0.33\,p=0.06$ \\
      & Full          & $\rho=0.26\,p=0.01$ & $\rho=0.29\,p=0.01$ \\
      & Isolated      & $\rho=0.16\,p=0.33$ & $\rho=0.01\,p=0.95$ \\
\hline
\end{tabular}
\caption{Results of the Spearman's test on the correlation between the rest-frame equivalent width and the kinematics, traced by the width in velocity space, of the \mgii\ and \civ\ absorbers and the Ly$\alpha$ luminosity of isolated galaxies and groups. In the case of multiple galaxies connected to the same absorber, we consider the full sample, the brightest, or the closest member only.}
\label{tab:GRPcorr}
\end{table}

Both the strength and the kinematics of the \civ\ absorbers show a tentative ($\approx2\sigma$ level) positive correlation ($\rho\gtrsim0.25$ and $p-$value$\lesssim0.06$) with the luminosity of galaxies in groups, without any significance dependence on the chosen sub-sample. On the other hand, the strength of the \mgii\ absorption is marginally sensitive ($\rho\gtrsim0.40$ and $p-$value$\lesssim0.06$) to the luminosity of the closest and the brightest component of groups, while no strong dependence is found ($p-$value$\gtrsim0.30$) for the kinematics of the cool gas. Finally, we observe that the \mgii\ equivalent width also positively correlates with the luminosity of isolated galaxies ($\rho=0.89\,p=0.02$). To better understand the underlying physical motivation for these results, we recall that a KS test yields a $p$-value$\approx 0.13,\,0.40,\,0.85$ ($p$-value$\approx 0.92,\,0.71,\,0.36$) when comparing the luminosity of isolated LAEs to that of the brightest, closest component and the full sample of groups, respectively, around the \mgii\ (\civ) absorbers. This is indicative of a high likelihood that the distributions of the two galaxy populations are similar and originate from the same parent one \citep[see also][]{Nielsen2018_grp}, consistent with our findings for LAEs around the \civ\ gas and the high degree of overlap between the two samples.

In light of this, since there is no statistical difference between the Ly$\alpha$ luminosity of isolated LAEs and groups, the possible dependence of the \mgii\ equivalent width on the luminosity of both the groups and the isolated galaxies can be interpreted as a hint that the absorption strength of the cool gas is sensitive to the galaxies' activity and the local environment. On the contrary, the more ionized \civ\ gas, which does not correlate significantly with isolated LAEs, seems more affected by the large-scale environment.



\section{Discussion}
\label{sec:discussion}

\subsection{Evolution of the covering fraction with redshift}
\label{sec:disc_CF_z}

\begin{figure*} 
\centering
\includegraphics[width=\textwidth]{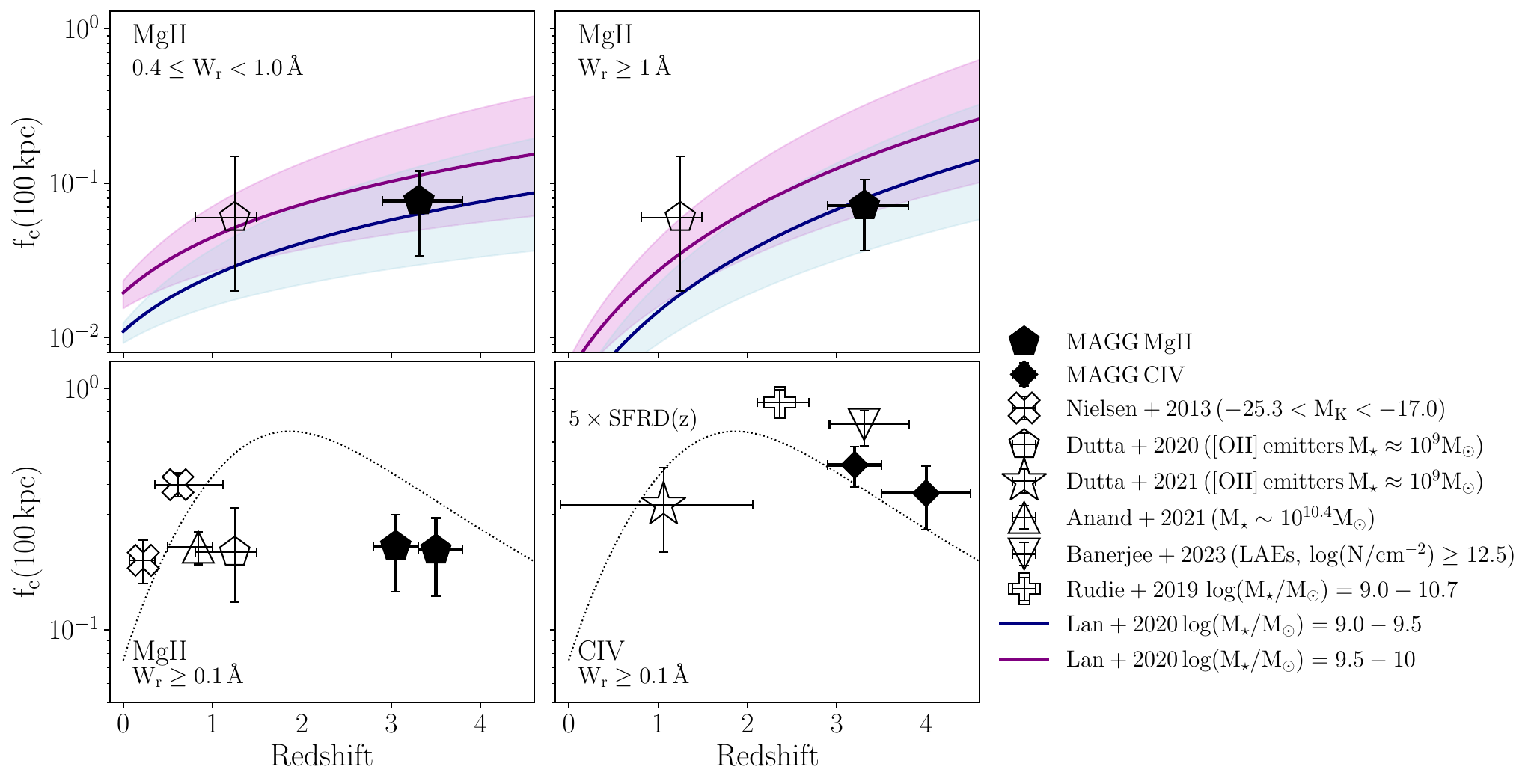}
\caption{\mgii\ and \civ\ cumulative covering fraction within $R<100\rm\,kpc$. Upper panels: The results for strong ($W_{\rm r}\geq1\,$\AA, left panel) and weak ($0.4\leq W_{\rm r}<1.0\,$\AA, right panel) \mgii\ absorbers observed around LAEs in MAGG (black) are consistent with the redshift evolution model calibrated at $z\lesssim 1.3$ by \citet{Lan2020} for star-forming galaxies with stellar mass $\log(M_\star/{\rm M_\odot})=9.0-9.5$ and $\log(M_\star/{\rm M_\odot})=9.5-10$. Also shown is the covering fraction of \mgii\ gas observed around [\ion{O}{III}] emitters by \citet{Dutta2020} at $z\approx1$. Lower panels: Results for $W_{\rm r}\geq0.1\,$\AA\ \mgii\ (right panel, black pentagon) and \civ\ (left panel, black diamond) samples, without any distinctions between weak and stronger absorbers. Also shown are the results from \citet{Nielsen2013b_general} (crosses), \citet{Dutta2020} (pentagon) and \citet{Anand2021} (up triangle, $W_{\rm r}\geq0.4\,$\AA), \citet{Dutta2021} (star), \citet{Rudie2019} (cross, $\log(N/\rm cm^{-2})\geq13.0)$ corresponding to $W_{r}\gtrsim0.04\,$\AA\ in linear regime), and \citet{Banerjee2023} (down triangle, $\log(N/\rm cm^{-2})\geq12.5)$ corresponding to $W_{r}\gtrsim0.01\,$\AA). Horizontal error bars mark the explored redshift range. The evolution of the \civ\ covering fraction is consistent with the star formation rate density as parametrized by \citet{Madau2014} (dotted line, arbitrary normalization). }
\label{fig:Disc:CF_with_z}
\end{figure*}

The distribution of the cool and metal-enriched gas traced by \mgii\ absorbers around star-forming galaxies has been extensively studied at $z\lesssim2$, but less so at times closer to the peak of the cosmic star formation rate density and in comparison to other gas phases, such as the more ionized gas traced by the \civ\ systems. The statistical sample of $z\approx3$ \mgii-LAEs systems surveyed in this work provides a key step forward in our understating of the evolution of the multiphase gas around galaxies at $z\gtrsim 3$. 

Many authors \citep[e.g.,][]{Turner2014,Rudie2019,Schroetter2021,Dutta2021} compared the incidence of different gas phases around star-forming galaxies at $z\lesssim2$, with \citet{Schroetter2021} and \citet{Dutta2021} directly focusing on the comparison between \mgii\ and \civ\ gas. The picture emerging from these studies is that the cool gas is concentrated at close separations from the galaxies, while the gas phase traced by higher ions is extended up to larger distances. However, this scenario is less clear with redshift, mostly due to the lack of large surveys of the \mgii-galaxies connection at $z\gtrsim2$. A first step in this direction comes from \citet{Lan2020}, who proposed an analytical model of the evolution of the \mgii\ covering fraction as a function of redshift and stellar mass. By calibrating the model with a sample of 15,000 pairs at $z\approx0.4-1.3$, they found that the covering fraction of strong ($W_{\rm r}\geq1\,$\AA) absorbers evolves as $(1+z)^{\alpha}$ with $\alpha\approx2$, following a trend that is similar to the evolution of the cosmic star formation rate density at $z\lesssim2$ \citep[see, e.g.,][]{Prochter2006, Matejek2012_FIRE, Zhu2013, Chen2017}. The evolution of weak ($W_{\rm r}=0.4-1\,$\AA) systems is less significant. However, several studies \citep{Steidel2010, Bordoloi2011, Chen2012, Schroetter2021, Dutta2020} did not find a significant redshift evolution when considering the contribution of weak and strong systems. 

The sample of $z\approx3-4$ \mgii\ lines presented in this work, along with the dense and complete set of LAEs identified in MAGG, offers an ideal dataset for testing the analytical predictions from \citet{Lan2020}, which we extrapolated from their results based on observations at $z\approx0.4-1.3$ to higher redshifts. To this end, we estimate the covering fraction of both strong and weak \mgii\ absorbers around LAEs within a projected separation $R\lesssim100\rm\,kpc$: $f_{\rm c}(100\rm\,kpc)$. In the following, we consider the entire sample of LAEs, and study in the next section \ref{sec:disc_zevolution_grp} the evolution of the covering fraction of different gas phases around group and isolated galaxies. 
In the upper panels in Figure \ref{fig:Disc:CF_with_z} we show that the $f_{\rm c}(100\rm\,kpc)$ estimated for the \mgii\ gas is consistent, within the uncertainties, with the predictions from \citet{Lan2020} for galaxies of stellar mass within the range $9.0<\log(M_\star/{\rm M_\odot})<9.5$ and $9.5<\log(M_\star/{\rm M_\odot})<10$, corresponding to typical stellar mass values expected for Ly$\alpha$ emitters \citep{Ouchi2020}. However, when incorporating the results from \citet{Dutta2020} at $z\approx1$, we find that these are consistent with the covering fraction estimated at $z\approx3$. While this could be in tension with the evolutionary trend suggested by \citet{Lan2020} especially for $W_{\rm r}\geq1\,$\AA, favouring a scenario where the cross-section of cool \mgii\ gas around galaxies remains relatively stable over cosmic time, it is important to note that large uncertainties in both observations and modelling prevent firm conclusions. 

Complementary to the analysis above, we consider the covering fraction of \mgii\ absorbers that are selected in equivalent width down to the completeness limit. We show in the lower-left panel of Figure \ref{fig:Disc:CF_with_z} the results for $W_{\rm r}\geq0.1\,$\AA\ \mgii\ lines, which corresponds to the full sample of absorbers connected to LAEs in MAGG, computed in two redshift bins centered at $z=3.05$ and $z=3.55$. To explore the time evolution, we compare to lower redshift observations \citep{Nielsen2013b_general, Dutta2020, Anand2021} obtained with the same threshold on the absorbers' equivalent width. The incidence of the cool gas is derived around star-forming galaxies with stellar masses of the order of $M_\star\approx10^9\rm M_\odot$ from \citet{Dutta2020} and $M_\star\approx10^{10.5}\rm M_\odot$ in all the other cases, which corresponds to $\approx0.5-1.0\rm\,dex$ above the typical mass of LAEs (except the sample of \citealp{Nielsen2013b_general} only characterized by an absolute K-Band magnitude, a proxy for galaxies' stellar mass, ranging between $-25.3<M_{\rm K}<-17.0$). This comparison shows no evolution across cosmic time, with $f_{\rm c}(100\rm\,kpc)\approx0.2$ at all redshifts. The current results do not rule out an evolution between $z\approx1.5-3$ due to lack of observations in this redshift range.

We next investigate how the cool and enriched gas distribution evolves as a function of redshift compared to the more ionized gas phase traced by \civ\ absorbers (lower panels in Figure \ref{fig:Disc:CF_with_z}). \citet{Schroetter2021} modeled the redshift evolution of the \civ\ covering fraction based on the connection of 19 \civ\ systems to [\ion{O}{II}] emitters at $z\approx1-1.5$ and did not find any significant evolution across the cosmic time. However, they observed that if, on the one hand, the size of the \mgii\ halo becomes progressively smaller with the cosmic time relative to that of the dark-matter halo, on the other hand, the evolution of \civ\ is more similar to that of the dark matter. At higher redshift, \citet{Banerjee2023} found that the covering fraction of \civ\ absorbers around LAEs, considering either the full sample or the isolated ones, does not evolve within a small redshift range between $z\approx2.9-3.8$ for \civ\ systems with $\log (N/{\rm cm^{-2}})\geq12.5$ (corresponding to $W_{\rm r}\gtrsim0.01\,$\AA\ in the linear regime). However, by combining the results from \maggv\ in two bins of redshift (centered at $z\approx3.2$ and $z\approx4$), to those at $z\approx1-2$ and $W_{\rm r}\geq0.1\,\text{\AA}$ from \citet{Dutta2021} and at $z\approx2$ from \citet{Rudie2019} (assuming $\log(N/\rm cm^{-2})\geq13.0$ which correspond to $W_{r}\gtrsim0.04\,$\AA\ in the linear regime) we observe that the \civ\ covering fraction increase with time up to $z\approx2$ and then decrease at $z\lesssim2$. 

In support of this result, \citet{Dutta2021} found that, within the virial radius, the covering fraction of the \civ\ gas increases by a factor of $\approx4$ from $z\lesssim1$ to $z\gtrsim1$, possibly driven by the redshift evolution of the gas density and the ionizing background or due to the galaxies at higher redshift having higher star formation rates. Indeed, we observe that the \civ\ covering fraction increases from the local Universe up to $z\approx2$, close to the peak of the cosmic star formation rate density \citep{Madau2014} (Figure \ref{fig:Disc:CF_with_z}), and then decreases. Conversely, the covering factor for the \mgii\ systems remains constant across cosmic time. As mentioned earlier, however, the large uncertainties in the observations allow the results at different redshifts to remain consistent, not ruling out a scenario where no evolution occurs. Moreover, additional uncertainty arises from the lack of a uniform equivalent width threshold across these studies, which is known to influence the measurement of the covering fraction (\maggv\ and \citealp{Banerjee2023}), and from the possibility that the samples could probe different galaxy populations at different redshifts.

Our observations suggest that the covering fraction of a cool and enriched gas phase around star-forming galaxies remains constant, within the uncertainties, throughout cosmic time, regardless of whether we consider strong or weak sub-samples of \mgii\ absorbers. The covering fraction of \civ\ absorbers seems to resemble more the cosmic star formation rate density evolution. Even larger surveys are needed in the future to reduce uncertainties and test against the scenario of no evolution, which is not yet completely ruled out.

\subsection{The effect of the large-scale galaxy environment across cosmic time}
\label{sec:disc_zevolution_grp}

\begin{figure*} 
\centering
\includegraphics[width=\textwidth]{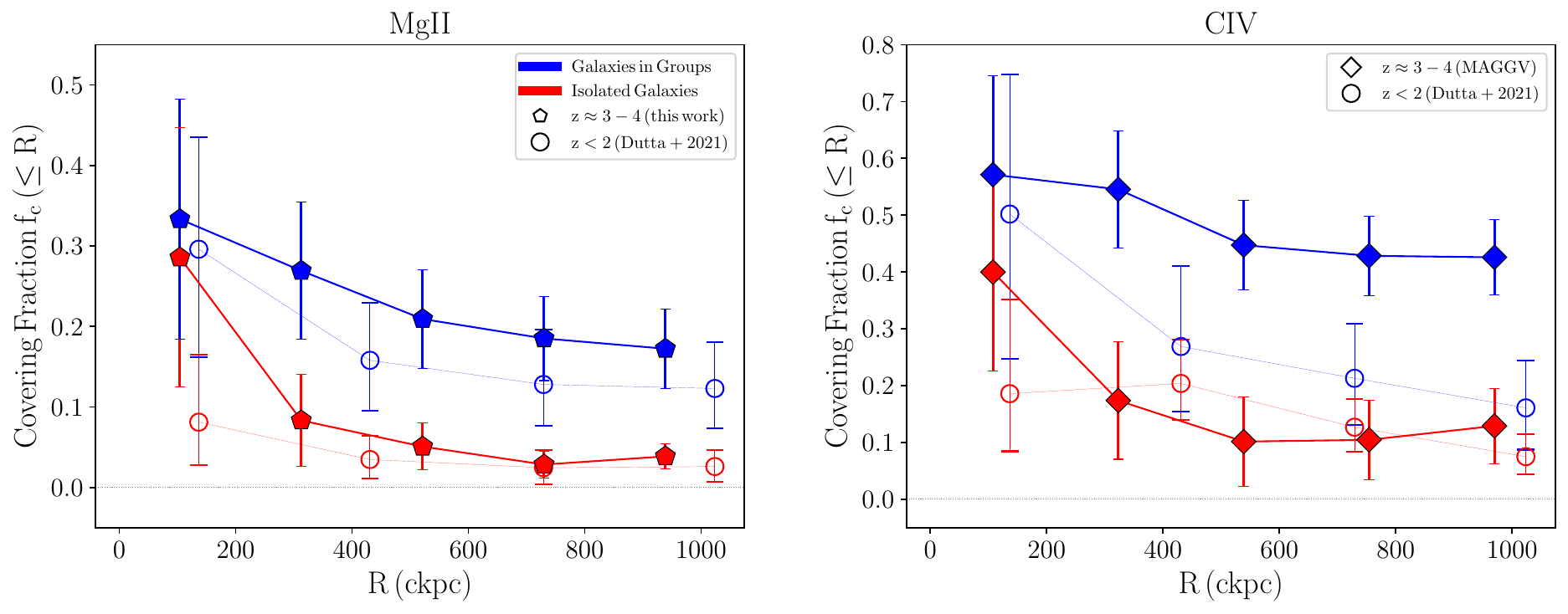}
\caption{Comparison of the covering fraction of \mgii\ (left panel) and \civ\ (right panel) absorbers with $W_{\rm r}\geq0.1\,$\AA\ detected around group and isolated galaxies in this work (filled points) with the results at $z\lesssim2$ from \citet{Dutta2021} (empty points). The distance has been converted into co-moving units at the mean redshift of the samples ($z=3.31$ in this work and $z=0.95$ for \citealp{Dutta2021}). The covering fraction is estimated for the closest LAE at $z\approx3-4$ and the most massive galaxy in groups at $z\lesssim2$. The scales on the y-axis are different in the two panels.}
\label{fig:Disc:CF_group_z}
\end{figure*}


We have shown in Figure \ref{fig:CFgroups} that, regardless the gas phase they are tracing, both the \mgii\ and the \civ\ absorption is enhanced around groups, relative to isolated LAEs, at any projected separation up to $R\approx250\rm\,kpc$. How does the large-scale galaxy environment affect the covering fraction and the distribution of different gas phases across cosmic time? To address this question, we now compare the results of Section~\ref{sec:effect_of_groups} with the findings from \citet{Dutta2021}, who studied the environmental effects on different gas phases at $z\lesssim2$ (see Figure \ref{fig:Disc:CF_group_z}). These authors found that $W_{\rm r}\geq0.1\,$\AA\ \mgii\ absorption-line systems are detected around groups with a probability $\approx2-3$ times higher compared to isolated galaxies, which is consistent with our findings at higher redshift. This result is proved to be valid even after matching the group and isolated samples in redshift, stellar mass and impact parameter, indicating that the environmental factors likely lead to the observed difference. Indeed, the \mgii\ covering fraction around the group and isolated LAEs at $z\approx3$ is consistent within $1\sigma$ uncertainties with the $z\lesssim2$ results around [\ion{O}{II}] emitters. Several mechanisms have been invoked at low redshift to explain this absorption excess. Modeling groups as a superposition of multiple haloes of isolated-like galaxies \citep[see, e.g.,][]{Bordoloi2011} does not always succeed in reproducing the observations \citep{Nielsen2018_grp, Dutta2020}, suggesting that effects like hydrodynamic and gravitational interactions or intragroup medium should not be neglected. Indeed, \citet{Dutta2021} interpreted the enhanced \mgii\ absorption in groups at $z\approx1$ as possibly due to stripping processes driven by the intra-group medium. Mergers and tidal interactions are also able to perturb the cool gas, especially at $z\gtrsim1$, when the occurrence of mergers is higher \citep[see, e.g.,][]{Conselice2008, Conselice2014}. However, we do not expect these processes to be equally efficient in the Universe at $z\approx3$, when filamentary structures have not viriliazed yet. Although the results shown in Section \ref{sec:env_abs_prop} seems to suggests that the \mgii\ gas in more sensitive to galaxies' individual halos in groups compared to the \civ, what is responsible for boosting the covering fraction of cool gas at high redshift and sustaining its abundance for several billions of years still remains an open question. 

A different picture emerges when considering the redshift evolution of the more ionized \civ\ gas in different galaxy environments. In \maggv, we observed that the \civ\ absorbers at $z\gtrsim3$ show a $\times3$ higher probability to be found around group than around isolated LAEs \citep[see also][]{Muzahid2021, Banerjee2023}. Conversely, at $z\lesssim2$, this gas is much less sensitive to the environment. Moreover, at $z\lesssim0.015$ \citet{Burchett2016} preferentially found $\log(N_\civ/\rm cm^{-2})>13.5$ \civ\ gas around galaxies in low-density regions and a lack of detections in rich and massive groups. Similar results have been found at $z\lesssim2$ by \citet{Dutta2021}, who showed that the incidence and the rest-frame equivalent width of the \civ\ absorbers increase only weakly with galaxy overdensity. In addition, they do not detect any significant difference comparing the \civ\ covering fraction of groups and isolated galaxies, with only a marginal enhancement around groups at $R\lesssim100\rm\,kpc$. In Figure \ref{fig:Disc:CF_group_z} (right panel), we show a direct comparison with the $W_{\rm r}\geq0.1\,$\AA\ \civ\ covering fraction around group and isolated galaxies at $z\gtrsim3$ in \maggv\ and $z\lesssim2$ by \citet{Dutta2021}. While the covering fraction around isolated galaxies does not vary significantly with time, that of \civ\ gas around group galaxies drops from $z\approx3$ to $z\lesssim2$.

In conclusion, we do not observe significant redshift evolution for the excess of cool \mgii\ gas around LAEs in groups, but the enhancement of the more ionized gas phase around group galaxies reduces at lower redshifts. As also argued by \citet{Burchett2016} and \citet{Dutta2021}, one interpretation of these results is that the \mgii\ absorbers arise from dense gas at small separation from the galaxies and are sensitive to the local environment, while the more diffuse and ionized gas phase is instead more extended (see details in Section \ref{sec:disc_EW_R}) and thus more sensitive to the conditions of the large-scale environments. Moreover, the low incidence rate of \civ\ absorbers found by both \citet{Burchett2016} and \citet{Dutta2021} in massive haloes with $M_{\rm h}\gtrsim10^{12}-10^{13}\rm\,M_\odot$ may be indicative of the majority of the \civ\ gas to be shifted to higher ionization states. Based on this, the effect of the large-scale galaxy environment on the ionization state of the gas can be responsible for the evolution of the covering fractions observed in Figure \ref{fig:Disc:CF_with_z} for the different gas phases.

\subsection{Radial distribution of different gas phases}
\label{sec:disc_EW_R}

\begin{figure} 
\centering
\includegraphics[width=\columnwidth]{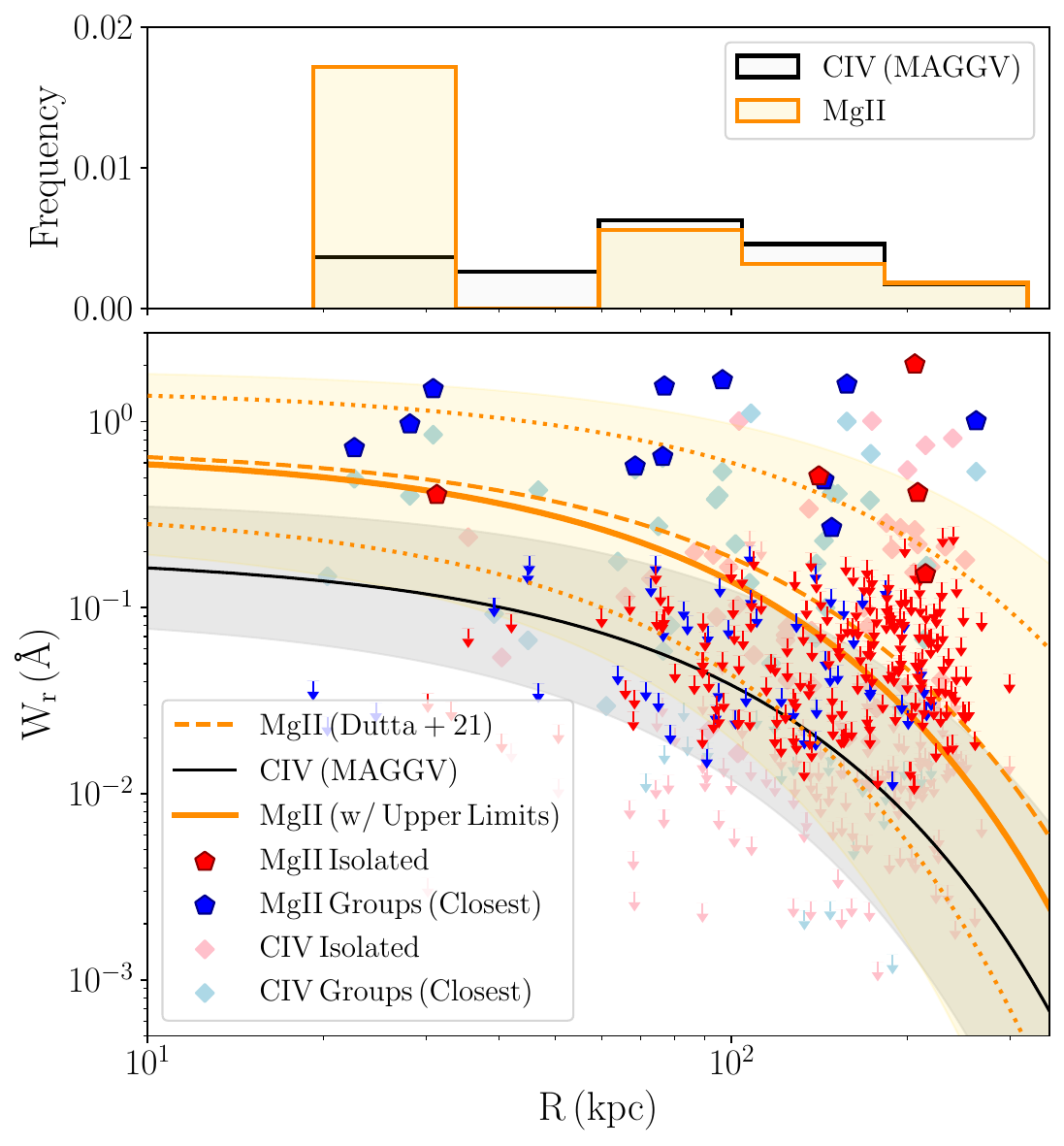}
\caption{Upper panel: Impact parameter distribution of LAEs around \mgii\ (orange) and \civ\ (black) absorbers. Only the member closest to the line of sight is considered for each group. Lower panel: Rest-frame equivalent width of the \mgii\ and \civ\ absorbers around the group (closest member, blue and light blue, respectively) and isolated LAEs (red and pink), as a function of the galaxies' projected separation. Arrows mark 3$\sigma$ upper limits where no absorption systems are detected around LAEs. The best log-linear models fitted to the data and the upper limits are shown as solid lines, while the shaded area marks the $1\sigma$ uncertainties. As a comparison, we show the \mgii\ best-fit log-linear model at $z\lesssim2$ from \citet{Dutta2021} (dashed orange line).}
\label{fig:Disc:EW_R}
\end{figure}

\begin{figure} 
\centering
\includegraphics[width=\columnwidth]{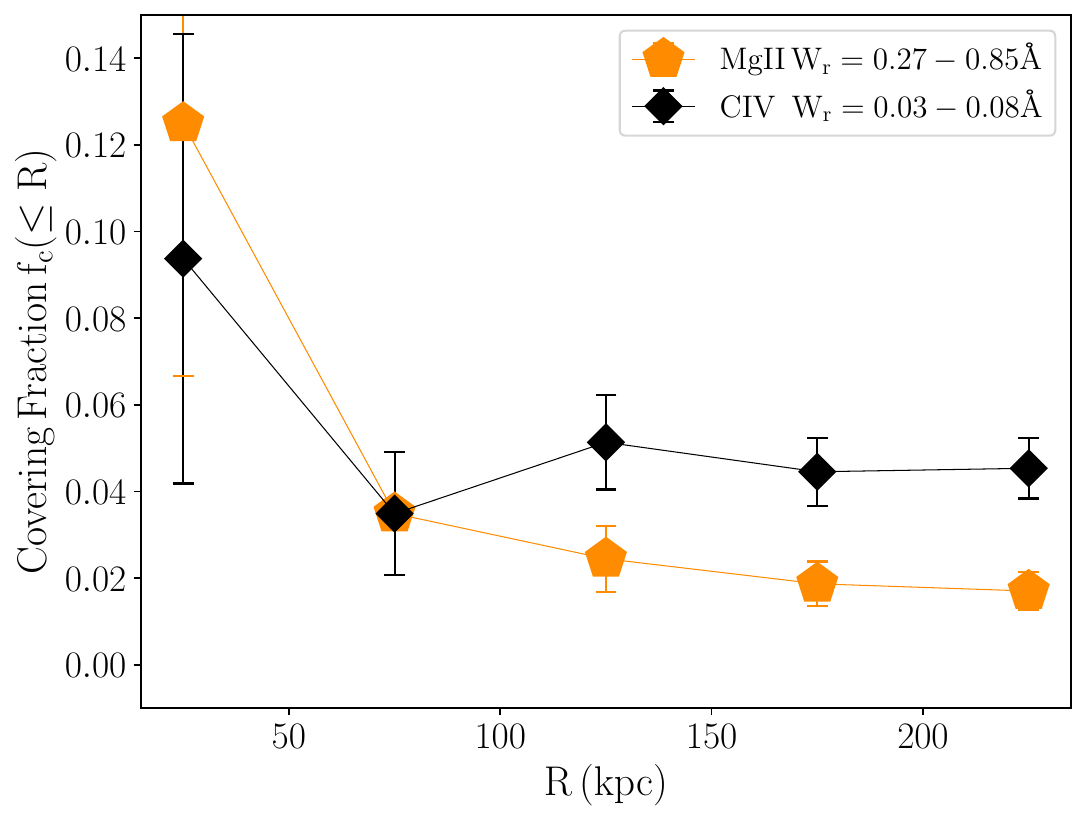}
\caption{Covering fraction of \mgii\ and \civ\ systems with $12.80<\log(N/\rm cm^{-2})\leq13.30$ (filled markers). The corresponding equivalent width values in the linear regime are reported in the legend. Within R$\leq100\rm\,kpc$, the covering fraction of the \mgii\ gas is similar to the \civ\, while at larger distances, we observe a significant excess of \civ\ absorbers and a decline in the \mgii\ covering fraction.}
\label{fig:Disc:CF_fixN}
\end{figure}

As introduced in Section \ref{sec:disc_zevolution_grp}, the radial distribution of the different gas phases around the galaxies provides insights into the physical mechanisms behind the evolution of the effect of the large-scale galaxy environment. Indeed, at low redshift, it is well established that the strength of the \mgii\ \citep[see e.g.][]{Lanzetta1990, Bergeron1991, Chen2010, Churchill2013, Nielsen2013b_general} and of the \civ\ \citep[see e.g.][]{Bordoloi2014,Burchett2016} absorption-line systems sharply decreases for increasing projected separations from the galaxies. The picture is complicated when considering the effect of the large-scale galaxy environment. At $z\lesssim2$ the \mgii\ rest-frame equivalent width drops beyond the galaxies' virial radius, but the relation flattens at large projected separations for \mgii\ and \civ\ systems around groups \citep{Chen2010, Bordoloi2011, Nielsen2018_grp, Dutta2020, Dutta2021, Nelson2021}. Overall, when the radial distribution of the cool gas is compared to that of the more ionized phase, the first seems to be concentrated at small distances from the associated galaxies, while the latter appears more extended and reaches projected separations several times larger than the galaxies' virial radius both in observations \citep[see e.g.][]{Dutta2021} and simulations \citep{Ford2013, Ho2020, Ho2021, Weng2024}. Moreover, \citet{Werk2014} found that the gas is ionized up to higher states with increasing distance from the host galaxy. How this condition evolves with redshift, especially in groups compared to isolated sources, may explain the evolution of the \civ\ covering fraction around multiple galaxies.

We have shown in \maggv\ that at $z\gtrsim3$ the \civ\ absorption strength does not decrease significantly for increasing projected distance from the associated LAEs and is indeed consistent with a flat radial profile \citep[see also][]{Muzahid2021, Banerjee2023}. Compared to the \civ\ gas, we show the equivalent width of the \mgii\ absorbers as a function of the projected distance from the closest component of each group (blue) and the isolated LAEs (red) in the bottom panel in Figure \ref{fig:Disc:EW_R}. We estimate $3\sigma$ upper limits on the absorbers' equivalent width in case any \mgii\ systems are detected around LAEs within $\pm500\rm\,km\,s^{-1}$. The radial profile of the absorption strength is modeled by a log-linear function following the method detailed in \citet{Dutta2020}. According to the best-fit parameters, $a=-0.061^{+0.188}_{-0.180}$ and $b=-0.001^{+0.001}_{-0.001}$, the radial profile of the \mgii\ absorbers shows a higher normalization compared to the \civ\ but decreases similarly at larger distances, while the strength of the detections remains flatter as a function of the projected separation. In addition, we do not observe any significant evolution across the cosmic time when comparing the \mgii\ equivalent width radial profile to the $z\approx1$ result from \citet{Dutta2020} and at $z\lesssim2$ from \citet{Dutta2021} (orange dashed line in Figure \ref{fig:Disc:EW_R}).

In the upper panel in Figure \ref{fig:Disc:EW_R}, we show the distributions of the impact parameters of LAEs connected to the \mgii\ (orange) and \civ\ absorbers (black) considering only the closest component of each group. To assess the statistical difference between the radial distribution of the two gas phases, we perform a KS test and obtain a $p-$value$\approx0.03$, which suggests that the two distributions are unlikely to arise from the same parent one. Combined with the excess of \mgii\ absorbers found at $R\lesssim40\rm\,kpc$, this result supports a picture in which different gas phases are possibly distributed differently around galaxies. We further explore this direction by measuring the \mgii\ and \civ\ covering fraction for two ranges of column density selected in the regime where the \mgii\ absorbers are not saturated (see Figure \ref{fig:EWoverlap}). This approach has the advantage of not being affected by the different selection functions of the two samples. 

Since the two different ions trace different gas densities at a given equivalent width, we control for the effect of weak \civ\ systems ($W_{\rm\civ}\lesssim0.1\,\text{\AA}$) incuded in the analysis above by measuring the covering fractions of absorbers stronger than a threshold on the column density. This is chosen so that the corresponding lower limit in equivalent width is higher than the completeness limit of the samples. In Figure \ref{fig:Disc:CF_fixN} we show the result for \mgii\ (orange) and \civ\ (black) absorbers with $12.80<\log(N/\rm cm^{-2})\leq13.30$ corresponding, in linear regime, to $W_{\rm\mgii}=0.26-0.85\,\text{\AA}$ and $W_{\rm\civ}=0.03-0.08\,\text{\AA}$ for the \mgii\ and \civ\, respectively. 
The \mgii\ gas is similar to the more ionized \civ\ at $R\leq100\rm\,kpc$ and steeply decreases at larger separations. On the contrary, the covering fraction of the \civ\ gas flattens at $R\geq100\rm\,kpc$ around a value $\approx3$ times larger than the \mgii. Altogether, this analysis provides evidence that the gas in a low ionization stage is more concentrated in close proximity to galaxies, while the more ionized gas phase traced by the \civ\ systems is more extended, consistently with the picture suggested at $z\lesssim2$. 


\section{Summary and conclusions}

In this paper, we have presented the results from a new 15.9h XSHOOTER program (PID 0109.A-0559; PI M. Galbiati), which complemented the MUSE Analysis of Gas around Galaxies (MAGG) survey with NIR spectroscopy of the background quasars. This dataset allowed us to study 47 \mgii\ and compare them to 220 \civ\ absorption line systems from \maggv\ as tracers of different gas phases around low-mass Ly$\alpha$ emitting galaxies at $z\approx3-4$ as a function of the large-scale galaxy environment. The main results are:

\begin{itemize}
\item We computed the Ly$\alpha$ luminosity function and the projected cross-correlation function of Ly$\alpha$ emitters (LAEs) identified within a velocity window of $|\Delta v|\leq500\rm\,km\,s^{-1}$ from the \mgii\ absorbers (Figure \ref{fig:LF_Xcorr}). LAEs cluster around the \mgii\ gas, as observed in previous MAGG analysis for other tracers (\civ, \hi). Compared to \civ, the luminosity function appears elevated in normalization, similar to that of strong \hi\ absorbers, and the projected cross-correlation function is better modeled by a steeper, $\gamma = 2.2$, power law. 
While \civ\ absorbers show a high rate of LAEs detections for large equivalent widths ($\gtrsim 80$ percent), the detection rate of LAEs near \mgii\ remains around $40~$percent, with a hint of decline towards the highest equivalent widths. An undetected (more dusty) galaxy population may be responsible for this difference. 
\item The large-scale galaxy environment shapes the covering fraction and the properties of both the \mgii\ and the \civ\ bearing gas. The probability of detecting the absorbers is a factor of $\approx3-4$ larger around LAEs in groups relative to isolated galaxies (Figure \ref{fig:CFgroups}). The strength, probed by the rest-frame equivalent width, and the absorbers' kinematics, traced by the width in velocity space, positively correlate with the number of associated galaxies (Figure \ref{fig:Prop_Ngal}). 
\item Extending the analysis to $z\lesssim 2$, we observe no significant evolution in the covering fraction of absorbers around LAEs when considering the entire galaxy sample (Figure \ref{fig:Disc:CF_with_z}). However, after splitting between groups and isolated LAEs (Figure \ref{fig:Disc:CF_group_z}), we observe 
the excess of \mgii\ absorbers around group galaxies persists from $z\approx3-4$ to $z\lesssim2$ without significant evolution. Conversely, the abundance of \civ\ gas around group LAEs seen at $z\approx3-4$ drops at $z\lesssim2$, 
matching the covering fraction observed around isolated galaxies.
\item The radial profile of the \mgii\ equivalent width resembles the one of  \civ\, up to distances of $\approx250\rm\,kpc$. 
However, when measuring the covering fraction above a fixed threshold of the column density, the more ionized \civ\ gas appears more extended than the cooler \mgii\ phase, which instead is more concentrated closer to the galaxies, as also observed at lower redshifts.
\end{itemize}

Complementing the MAGG survey with NIR quasar spectroscopy allows us to trace simultaneously different gas phases around $z\approx3$ galaxies and to extend the study of the cosmic gas to the surroundings of the low-mass end of the galaxies' mass function. 
The fundamental statistics (overdensity, luminosity function, and cross-correlation functions) of LAEs in regions selected by metals in different ionization stages (\mgii, \civ\ and \siv\ absorbers) and by neutral hydrogen share strong similarities, implying that galaxies are distributed inside cosmic structures that contain multiphase gas in composition and temperature. Yet, our analysis has also uncovered differences in the clustering of LAEs, allowing us to paint a more refined picture of the location of each gas component relative to galaxies. In MAGG~V, by comparing the distributions of \hi\ and \civ\ absorbers, we concluded that LAEs lie along gas filaments that contain the majority of the optically thick hydrogen seen at $z\approx 3$, as well as enriched gas traced by the strong \civ\ systems. Lower equivalent width absorbers are believed to trace instead a more diffuse gas phase extended around LAEs. \mgii\ gas shares similarities with the strong \civ\ absorbers and the \hi\ gas, but with a steeper projected correlation function that suggests that this cool and enriched phase is more centrally concentrated near LAEs. However, with a $40$ percent detection rate of LAEs for strong \mgii\ absorbers, an additional (star-forming and dusty) galaxy population is likely missing from our MUSE survey. 

 The MAGG survey has also added the dimension of the large-scale galaxy environment. Combining our previous results and this analysis, we find that the number of galaxies associated with the absorbers can shape their strength and kinematics for the \mgii\ and \civ\ gas. The covering fraction of these metal absorbers around group or isolated galaxies is the quantity that differs the most among the various gas phases. While the gravitational and hydrodynamic mechanisms able to enhance the absorbers' cross-section or displace the enriched gas from the inner part of galaxies are better known at $z\lesssim2$, the processes responsible for the excess of \mgii\ and \civ\ bearing gas observed at higher redshifts remain currently less clear. For instance, \citet{Muzahid2021} (see also \maggv\ and \citealp{Banerjee2023}) proposed that the abundance of gas around galaxies in groups might be due to the large-scale structure in which groups themselves are embedded. 

Although both the \mgii\ and the \civ\ absorbers are found to be a factor $\approx3-4$ times more abundant around groups of LAEs, we find a substantial difference in the distribution of the two gas phases, with a large fraction of \mgii\ absorbers at close separations from the galaxies compared to the \civ\, that is instead tracing a more ionized gas phase that is likely extended up to larger distances from the associated LAEs. This goes with the picture painted by \citet{Werk2014} at $z\approx0.2$ and \citet{Dutta2021} at $z\approx1$, according to which the gas is ionized to higher stages with increasing distance from the galaxies. This difference in the distribution and ionization state of the gas might drive the sharp decrease that we observe for the \civ\ covering fraction in groups from $z\approx3$ to $z\lesssim2$. In this case, the lack of difference in the covering fraction of \civ\ absorbers around lower-redshifts groups finds a possible explanation if the gas is ionized to higher stages and is more sensitive to the large-scale, rather than the local, environment. 

Further progress in completing our view of the multiphase gas near star-forming galaxies at cosmic noon requires extending our view of the galaxy population into the infrared and sub-mm to select via optical or molecular lines more massive and highly star-forming galaxies that are not easily detected in the UV rest-frame. 
In parallel, a thorough exploration of the scenario painted above in cosmological simulations is required, to also learn about plausible mechanisms at the origin of the differences uncovered in group and more isolated galaxies.

\begin{acknowledgements}
We thank the anonymous referee for the careful reading of this paper and the suggestions which improved the manuscript. This project has received funding from the European Research Council (ERC) under the European Union's Horizon 2020 research and innovation programme (grant agreement No 757535) and by Fondazione Cariplo, grant No 2018-2329. SC gratefully acknowledges support from the European Research Council (ERC) under the European Union’s Horizon 2020 research and innovation programme grant agreement no 864361. 
This work is based on observations collected at the European Organisation for Astronomical Research in the Southern Hemisphere under ESO programme IDs 197.A-0384, 065.O-0299, 067.A-0022, 068.A-0461, 068.A-0492, 068.A-0600, 068.B-0115, 069.A-0613, 071.A-0067, 071.A-0114, 073.A-0071, 073.A-0653, 073.B-0787, 074.A-0306, 075.A-0464, 077.A-0166, 080.A-0482, 083.A-0042, 091.A-0833, 092.A-0011, 093.A-0575, 094.A-0280, 094.A-0131, 094.A-0585, 095.A-0200, 096.A-0937, 097.A-0089, 099.A-0159, 166.A-0106, 189.A-0424, 0109.A-0559.
This work used the DiRAC@Durham facility managed by the Institute for Computational Cosmology on behalf of the STFC DiRAC HPC Facility (\url{www.dirac.ac.uk}). The equipment was funded by BEIS capital funding via STFC capital grants ST/P002293/1, ST/R002371/1 and ST/S002502/1, Durham University and STFC operations grant ST/R000832/1. DiRAC is part of the National e-Infrastructure. This research made use of Astropy \citep{Astropy2013}.
\end{acknowledgements}


\bibliographystyle{aa}
\bibliography{Reference_v1}


\begin{appendix}

\section{Kinematics}
\label{app:kinematics}

\begin{figure} 
\centering
\includegraphics[width=\columnwidth]{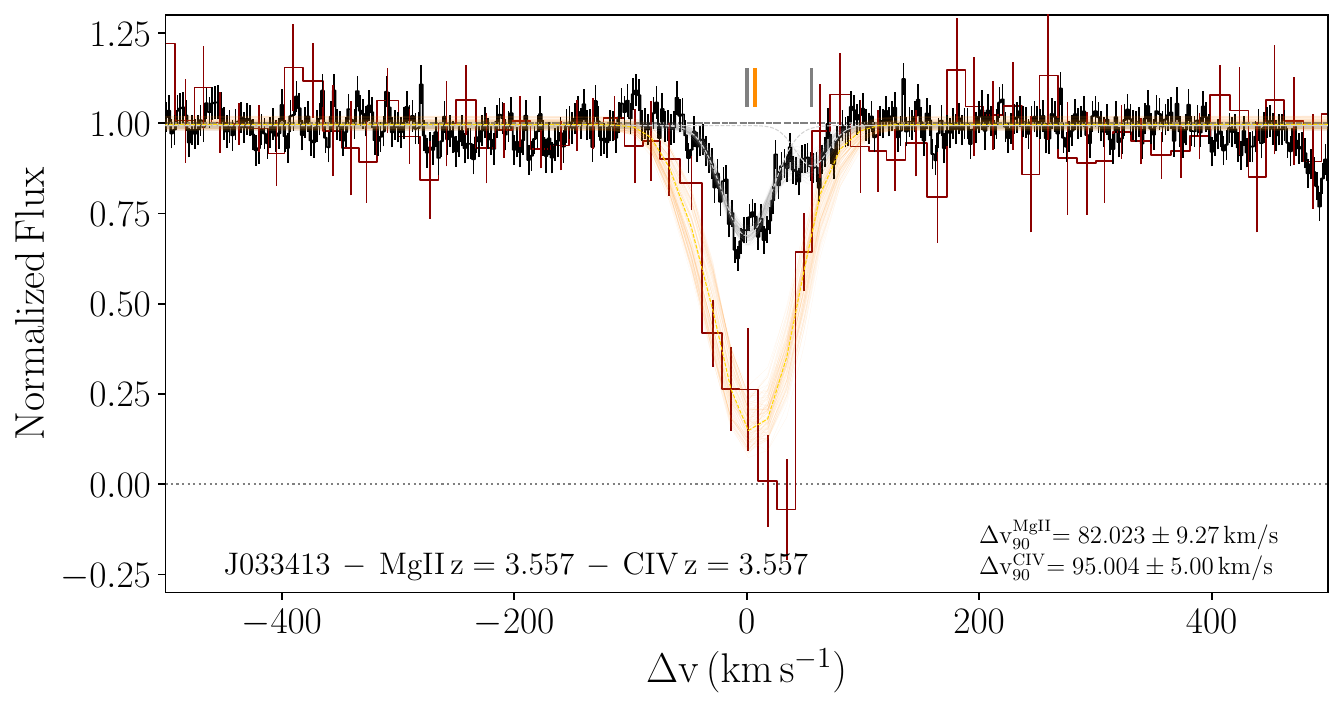}
\includegraphics[width=\columnwidth]{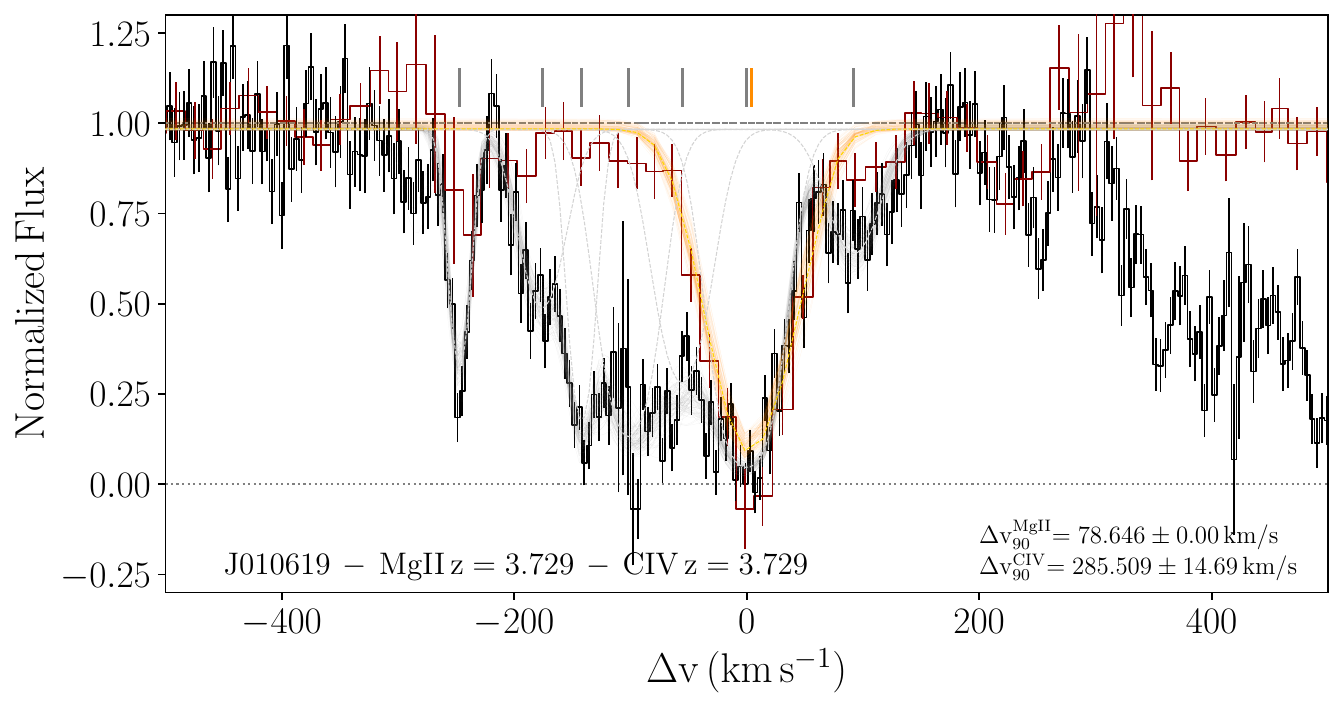}
\caption{Examples of \mgii\ (dark-red spectrum and orange best-fit profile) and \civ\ (black spectrum and gray profile) absorbers matched in velocity space. $\Delta v=0\rm\,km\,s^{-1}$ corresponds to the redshift of the \civ\ component with the largest column density. We show in the upper panel a representation of the case $\Delta v_{90}^{\civ}\approx\Delta v_{90}^{\mgii}$ (matched system) and in the lower panel the case $\Delta v_{90}^{\civ}>2\times\Delta v_{90}^{\mgii}$ and with only a partial alignment between the two ions (mis-matched system). The sightline, redshifts and measure of the velocity width are reported in the figures.}
\label{fig:app_kinematics_examples}
\end{figure}

\begin{figure} 
\centering
\includegraphics[width=\columnwidth]{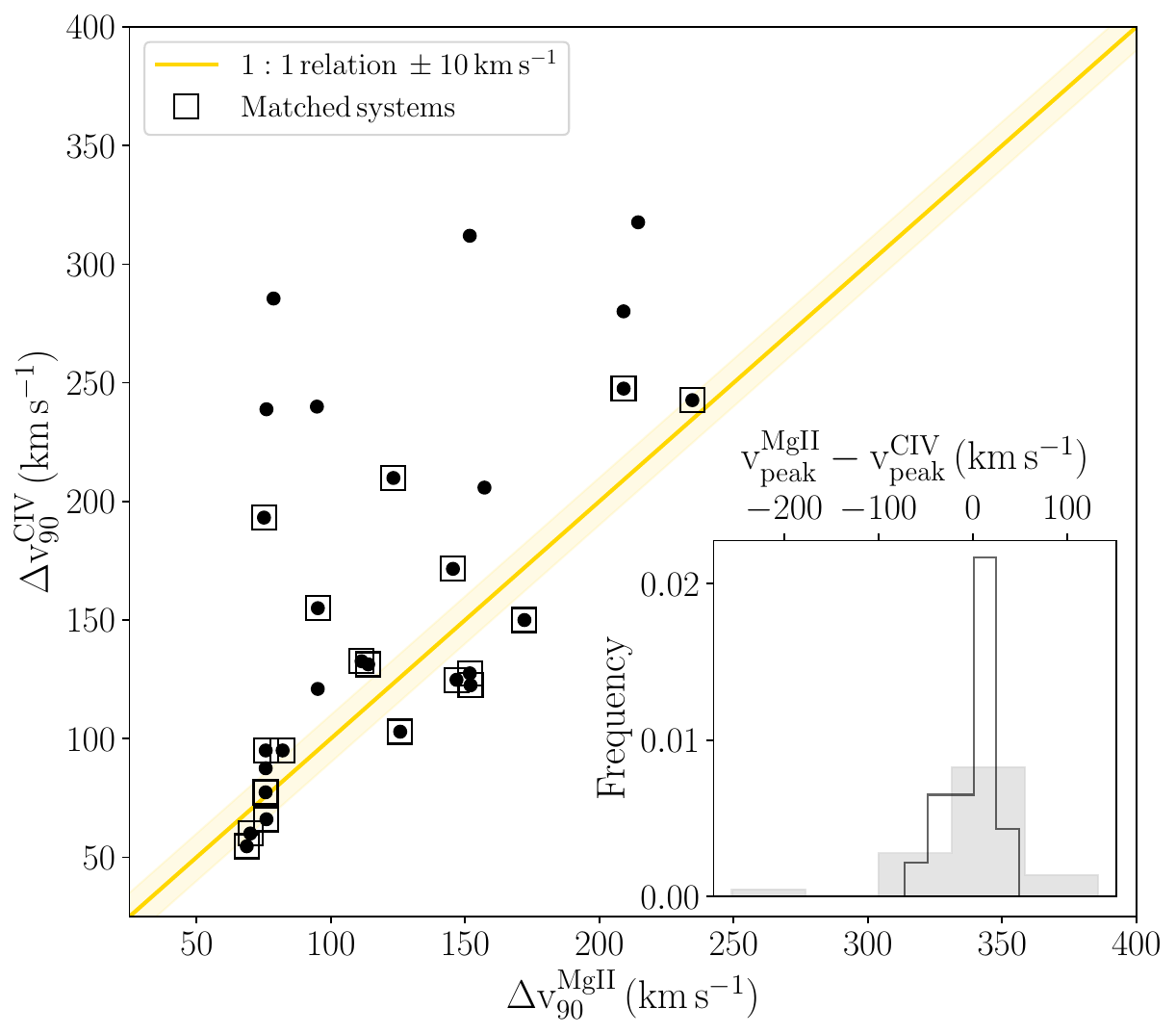}
\caption{Comparison between the velocity width of the \mgii\ (x-axis) and the \civ\ (y-axis) systems found to be matched within $\pm150\rm\,km\,s^{-1}$ along the line-of-sight. We show the $\Delta v_{90}^\mgii=\Delta v_{90}^\civ$ relation (gold line) with the shaded area marking the typical $1\sigma$ uncertainty on the kinematics. Open squares highlight the \mgii\ and \civ\ systems whose absorption line profiles are matched in velocity space. The inset shows the velocity shift between the peak of the \mgii\ optical depth relative to the \civ\ for the full sample (lightgray) and restricted to only the matched systems (black).}
\label{fig:app_kinematics}
\end{figure}

As detailed in Section \ref{sec:overlap_MAGG}, we found 28 \mgii\ absorbers matching \civ\ lines from \maggv\ in velocity space within $\pm150\rm\,km\,s^{-1}$. Among these, we identify 19 cases in which the \mgii\ and \civ\ absorption line profiles is matched in velocity space (see an example in the upper panel in Figure \ref{fig:app_kinematics_examples}), while the remaining systems are mis-macthed and only partially aligned (lower panel in Figure \ref{fig:app_kinematics_examples}). The former systems are likely arising from the same gas clouds and are thus direct evidence of the multiphase nature of the CGM of Ly$\alpha$ emitting galaxies. \citet{Rudie2019} probed the gas clouds in star-forming galaxies' CGM by tracing a variety absorbers, from low to high ions, at $z\approx3$. They observed different kinematics for the elements in different ionization stages: indeed, \ion{O}{VI} and \civ\ systems show broader absorption features compared to singly, doubly and triply ionized species. These aligned systems identified in MAGG allow us to extend this analysis at $z\gtrsim3$. However, the different resolution of the optical and NIR quasar spectra in which we identified the \civ\ and \mgii\ absorbers, respectively, prevents from directly compare the individual Voigt components. We thus show in Figure \ref{fig:app_kinematics} an alternative tracer of the gas kinematics, i.e. the width of the lines in velocity space, for the \mgii\ absorbers (x-axis) relative to that of the matching \civ\ lines (y-axis). We find that a large fraction of the data-points are scattered around the one-to-one relation, corresponding to $\Delta v_{90}^{\civ}=\Delta v_{90}^{\mgii}$. A few systems show kinematics that is consistent within $1\sigma=\pm10\rm\,km\,s^{-1}$, the typical uncertainty on the measurements of the velocity width for the two ions. In more details, $\approx71\%$ ($\approx58\%$) of the \civ\ systems (matched \civ\ systems) show, on average, a $\approx1.2$ times larger $\Delta v_{90}$ and is scattered around $\Delta v_{90}^{\civ}=\Delta v_{90}^{MgII}\pm2.5\sigma$ (see an example in the upper panel in Figure \ref{fig:app_kinematics_examples}). About 30 percent of the systems populates the upper-left corner of Figure \ref{fig:app_kinematics}, corresponding to the region of $\Delta v_{90}^{\civ}>2\times\Delta v_{90}^{\mgii}$. In these cases the \mgii\ absorption profile significantly differs from that of the \civ\ and is partially aligned (see the lower panel in Figure \ref{fig:app_kinematics_examples}). Indeed, the mild difference in the kinematics is washed out when only the matched systems are taken into account.

In the inset in Figure \ref{fig:app_kinematics} we show a complementary tracer of the kinematics of the two ions: the velocity shift between the peak of the \mgii\ optical depth relative to that of the \civ. The distribution (black histogram for the aligned \mgii\ and \civ\ systems) is centered around a median value of $v_{\rm peak}^{\rm\mgii}-v_{\rm peak}^{\rm\civ}\approx2.7\rm\,km\,s^{-1}$, which is indicative of only a mild difference between the \mgii\ and \civ\ kinematics and consistent with the analysis above. Including also the partially aligned systems (lightgray histogram) results in a median value $v_{\rm peak}^{\rm\mgii}-v_{\rm peak}^{\rm\civ}\approx5.4\rm\,km\,s^{-1}$) and does not change the outcome.

In conclusion, $\approx58\%$ of the \mgii-\civ\ systems matched along the line-of-sight shows a velocity width that is consistent within $\approx2.5\sigma$. While in mis-macthed systems we typically observe $\Delta v_{90}^{\civ}\gtrsim2\times\Delta v_{90}^{\mgii}$, we do not find any \mgii\ absorber significantly broader than the corresponding \civ\ line.

\end{appendix}


\end{document}